\pdfoutput=1
\documentclass[aps,prd,amsmath,floats,floatfix, twocolumn,
superscriptaddress,nofootinbib,showpacs,longbibliography]{revtex4-1}

\usepackage[T1]{fontenc}
\usepackage[utf8]{inputenc}
\usepackage{lmodern}

\usepackage{verbatim}

\usepackage[dvipsnames, usenames]{xcolor}
\definecolor{linkcolor}{rgb}{0.0,0.3,0.5}
\usepackage[hypertexnames=false, unicode, colorlinks=true, linkcolor=linkcolor,
citecolor=linkcolor, filecolor=linkcolor,urlcolor=linkcolor,
pdfusetitle]{hyperref}

\usepackage[all]{hypcap}
\usepackage{graphicx}
\usepackage{xspace}
\usepackage{amssymb}
\usepackage[normalem]{ulem}
\usepackage{bm}

\usepackage{blindtext}

\usepackage{microtype}

\usepackage[]{subcaption}

\graphicspath{%
  {figs/}%
}

\DeclareMathAlphabet{\mathpzc}{OT1}{pzc}{m}{it}

\newcommand{\roughly}{\mathchar"5218\relax\,}
\newcommand{\into}{\!\times\!\relax}

\newcommand{\h}{\mathpzc{h}}
\newcommand{\hCopr}{\mathpzc{h}^{\mathrm{copr}}}
\newcommand{\hCoorb}{\mathpzc{h}^{\mathrm{coorb}}}
\newcommand{\hlm}{\mathpzc{h}_{\ell m}}
\newcommand{\hlmCopr}{\mathpzc{h}^{\mathrm{copr}}_{\ell m}}
\newcommand{\hlmCoorb}{\mathpzc{h}^{\mathrm{coorb}}_{\ell m}}
\newcommand{\OmegaCoorb}{\Omega^{\mathrm{coorb}}}

\newcommand{\chieffCoorb}{\chi^{\mathrm{coorb}}_{\mathrm{eff}}}

\newcommand{\bchi}{\bm{\chi}}
\newcommand{\bchiCoorb}{\bm{\chi}^{\mathrm{coorb}}}
\newcommand{\chiCoorb}{\chi^{\mathrm{coorb}}}
\newcommand{\chihatCoorb}{\hat{\chi}^{\mathrm{coorb}}}
\newcommand{\bchidotCoorb}{\dot{\bm{\chi}}^{\mathrm{coorb}}}
\newcommand{\bv}{\bm{v}}
\newcommand{\bL}{\bm{L}}
\newcommand{\bLambda}{\bm{\Lambda}}
\newcommand{\tDyn}{t^{\mathrm{dyn}}}
\newcommand{\tCoorb}{t^{\mathrm{coorb}}}
\newcommand{\tmHundred}{t\!=\!-100M}
\newcommand{\Q}{\hat{Q}}

\newcommand{\eobmodel}{SEOBNRv3\xspace}

\newcommand{\RemnantModel}{NRSur7dq4Remnant\xspace}

\newcommand\caltech{\affiliation{TAPIR 350-17, California Institute of
Technology, 1200 E California Boulevard, Pasadena, CA 91125, USA}}
\newcommand{\AEI}{\affiliation{Max Planck Institute for Gravitational Physics
    (Albert Einstein Institute), Am M\"uhlenberg 1, Potsdam 14476, Germany}} %
\newcommand{\Cornell}{\affiliation{Center for Radiophysics and Space
    Research, Cornell University, Ithaca, New York 14853, USA}}
\newcommand{\UMassD}{\affiliation{Department of Mathematics,
    Center for Scientific Computing and Visualization Research,
    University of Massachusetts, Dartmouth, MA 02747, USA}}
\newcommand\olemiss{\affiliation{Department of Physics and Astronomy,
The University of Mississippi, University, MS 38677, USA}}
\newcommand{\bham}{\affiliation{School of Physics and Astronomy and Institute
for Gravitational Wave Astronomy, University of Birmingham, Birmingham, B15
2TT, UK}}

\begin{document}

\title{Surrogate models for precessing binary black hole simulations with
unequal masses}

\author{Vijay Varma}
\email{vvarma@caltech.edu}
\caltech

\author{Scott E. Field}
\UMassD

\author{Mark A. Scheel}
\caltech

\author{Jonathan Blackman}
\caltech

\author{Davide Gerosa}
\bham

\author{Leo C. Stein}
\olemiss

\author{Lawrence E. Kidder}
\Cornell

\author{Harald P. Pfeiffer}
\AEI

\hypersetup{pdfauthor={Varma et al.}}

\date{\today}

\begin{abstract}
Only numerical relativity simulations can capture the full complexities of
binary black hole mergers. These simulations, however, are prohibitively
expensive for direct data analysis applications such as parameter estimation.
We present two new fast and accurate surrogate models for the outputs of these
simulations: the first model, NRSur7dq4, predicts the gravitational waveform
and the second model, \RemnantModel, predicts the properties of the remnant
black hole. These models extend previous 7-dimensional, non-eccentric
precessing models to higher mass ratios, and have been trained against 1528
simulations with mass ratios $q\leq4$ and spin magnitudes $\chi_1,\chi_2 \leq
0.8$, with generic spin directions. The waveform model, NRSur7dq4, which begins
about 20 orbits before merger, includes all $\ell \leq 4$ spin-weighted
spherical harmonic modes, as well as the precession frame dynamics and spin
evolution of the black holes.  The final black hole model, \RemnantModel,
models the mass, spin, and recoil kick velocity of the remnant black hole. In
their training parameter range, both models are shown to be more accurate than
existing models by at least an order of magnitude, with errors comparable to
the estimated errors in the numerical relativity simulations. We also show that
the surrogate models work well even when extrapolated outside their training
parameter space range, up to mass ratios $q=6$.
\end{abstract}

\maketitle

\section{Introduction}
\label{sec:introduction}

As the LIGO~\cite{TheLIGOScientific:2014jea} and Virgo~\cite{TheVirgo:2014hva}
detectors reach their design sensitivity, gravitational wave (GW)
detections~\cite{Abbott:2016blz, TheLIGOScientific:2017qsa, Abbott:2016nmj,
Abbott:2017vtc, Abbott:2017gyy, Abbott:2017oio, LIGOScientific:2018mvr} are
becoming routine~\cite{Aasi:2013wya, LIGOScientific:2018jsj}. To maximize the
science output of the data collected by the network of detectors, it is crucial
to accurately model the source of the GWs.  Among the most important sources
for these detectors are binary black hole (BBH) systems, in which two black
holes (BHs) lose energy through GWs, causing them to inspiral and eventually
merge.

Numerical relativity (NR) simulations are necessary to accurately model the
late inspiral and merger stages of the BBH evolution.  These simulations
accurately solve Einstein's equations to predict the evolution of the BBH
spacetime. The most important outputs of NR simulations are the gravitational
waveform and the mass, spin, and recoil kick velocity of the remnant BH left
after the merger.

For interpreting detected signals, model waveforms are used to compare with
detector data and infer the properties of the source~\cite{CutlerFlanagan1994,
TheLIGOScientific:2016wfe, Veitch:2014wba}. The mass and spin of the remnant
determine the black hole ringdown frequencies, which are used in testing
general relativity~\cite{TheLIGOScientific:2016src, LIGOScientific:2019fpa,
Ghosh:2017gfp}. In addition, the recoil kick is astrophysically important
because it can cause the remnant BH to be ejected from its host
galaxy~\cite{Campanelli:2007cga, Gonzalez:2007hi, Gerosa:2014gja}.

Unfortunately, NR simulations are too expensive to be directly used in data
analysis applications and incorporated into astrophysical models. As a result,
several approximate models that are much faster to evaluate have been developed
for both waveforms~\cite{Khan:2018fmp, Cotesta:2018fcv, London:2017bcn,
    Pan:2013rra, Bohe:2016gbl, Khan:2015jqa, Hannam:2013oca,
    Taracchini:2013rva, Pan:2011gk, Mehta:2017jpq, Babak:2016tgq} and remnant
    properties~\cite{Hofmann:2016yih, Barausse:2012qz, Jimenez-Forteza:2016oae,
        Healy:2016lce, Healy:2014yta, Gonzalez:2006md, Campanelli:2007ew,
        Lousto:2007db, Lousto:2012su, Lousto:2012gt, Gerosa:2016sys,
        Healy:2018swt, Herrmann:2007ex, Campanelli:2007cga, Gonzalez:2007hi,
    Rezzolla:2007rz, Rezzolla:2007rd, Kesden:2008ga, Tichy:2008du,
Barausse:2009uz, Zlochower:2015wga}. These models typically assume an
underlying phenomenology based on physical motivations, and calibrate any
remaining free parameters to NR simulations.

Among BBHs, systems with BH spins that are misaligned with respect to the
orbital angular momentum are complicated to model analytically or
semi-analytically. For these systems, the spins interact with both the orbital
angular momentum and each other, causing the system to precess about the
direction of the total angular momentum~\cite{Apostolatos:1994pre}. This
precession is imprinted on the waveform as characteristic modulations in the
amplitude and frequency of the GWs, and can be used to extract information
about the spins of the source. One important application of the extracted spins
is to distinguish between formation channels of BBHs~\cite{Gerosa:2013laa,
Vitale:2015tea, Farr:2017gtv, Gerosa:2018wbw}.

The precessing BBH problem for quasicircular orbits is parametrized by seven
parameters: the mass ratio $q=m_1/m_2\geq1$ and two spin vectors $\bchi_{1,2}$,
where the index 1 (2) refers to the heavier (lighter) BH. The total mass scales
out of the problem and does not constitute an additional parameter for
modeling.  The surrogate models of Ref.~\cite{Blackman:2017pcm} for the
gravitational waveform, and Ref.~\cite{Varma:2018aht} for the remnant
properties, were the first to model the $7-$dimensional space of generically
precessing BBH systems, albeit restricted to mass ratios $q\leq 2$, and
dimensionless spin magnitudes $\chi_{1,2} \leq 0.8$. Trained directly against
numerical simulations, these models do not need to introduce additional
assumptions about the underlying phenomenology of the waveform or remnant
properties that necessarily introduces some systematic error. Through
cross-validation studies, it was shown that both these models achieve
accuracies comparable to the numerical simulations
themselves~\cite{Blackman:2017pcm, Varma:2018aht}, and as a result, are the
most accurate models currently available for precessing systems, within their
parameter space of validity.

In this paper, we present extensions of the above surrogate models to larger
mass ratios. Our new surrogate models are called NRSur7dq4 and \RemnantModel,
for the gravitational waveform and remnant properties, respectively. They are
trained against 1528 precessing NR simulations with mass ratios $q\leq4$, spin
magnitudes $\chi_1,\chi_2\leq 0.8$, and generic spin directions. Both models
are made publicly available through the gwsurrogate~\cite{gwsurrogate} and
surfinBH~\cite{surfinBH} Python packages; example evaluation codes are provided
at Ref.~\cite{SpECSurrogates} and Ref.~\cite{surfinBH}, respectively, for
NRSur7dq4 and \RemnantModel.

The rest of the paper is organized as follows. Section.~\ref{sec:prelims}
covers some preliminaries to set up the modeling problem for precessing BBH
systems.  Section~\ref{sec:nr_simulations} describes the training simulations.
Sec.~\ref{sec:waveform_sur} describes the NRSur7dq4 waveform surrogate model.
Section~\ref{sec:remnant_sur} describes the \RemnantModel remnant properties
surrogate model. Section~\ref{sec:results} compares these models against NR
simulations to assess their accuracy. Finally, Sec.~\ref{sec:conclusion}
presents some concluding remarks.  In App.~\ref{sec:extrap} we examine how
accurate these models are when extrapolated beyond mass ratio $q=4$, and in
App.~\ref{sec:PBandJ} we investigate some features in the error distribution of
the NR simulations.

\section{Preliminaries and notation}
\label{sec:prelims}

It is convenient to combine the two polarizations of the waveform into a single
complex, dimensionless strain  $\h = h_{+} -i h_{\times}$, and to represent the
waveform on a sphere as a sum of spin-weighted spherical harmonic modes:
\begin{equation}
    \h(t, \iota, \varphi_0) = \sum^{\infty}_{\ell=2} \sum_{m=-l}^{l}
        \hlm(t) ~_{-2}Y_{\ell m}(\iota, \varphi_0).
\label{eq:spherical_harm}
\end{equation}
Here $_{-2}Y_{\ell m}$ are the spin$\,=\!\!-2$ weighted spherical harmonics,
and $\iota$ and $\varphi_0$ are the polar and azimuthal angles on the sky in
the source frame.

For nonprecessing systems, the direction of orbital angular momentum ($\bL$) is
fixed and the $\hat{z}$ direction of the source frame is chosen to be along
$\hat{\bL}$ by convention. The gravitational radiation is strongest along the
directions parallel and antiparallel to $\hat{\bL}$. Therefore, for
nonprecessing systems the quadrupole modes ($\ell=2, m=\pm 2$) dominate the sum
in Eq.~(\ref{eq:spherical_harm}), but the nonquadrupole modes can become
important at large mass ratios or $\iota$ close to $\pi/2$~\cite{Varma:2016dnf,
    Capano:2013raa, Littenberg:2012uj, Bustillo:2016gid, Brown:2012nn,
Varma:2014jxa, Graff:2015bba, Harry:2017weg, Bustillo:2015qty,
Pekowsky:2012sr}.

\begin{figure}[thb]
\includegraphics[width=0.5\textwidth]{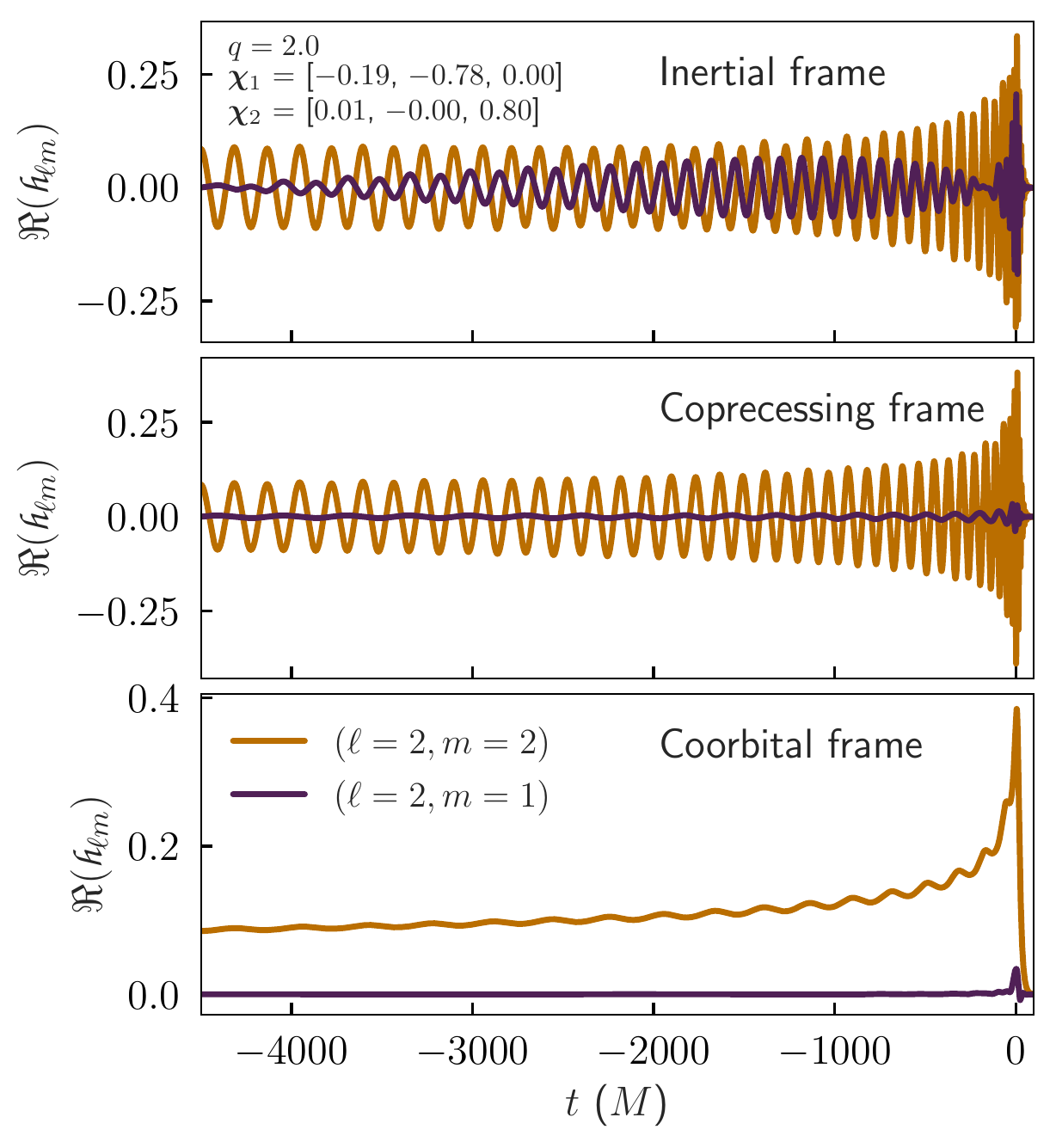}
\caption{The real part of the $(2,2)$ and $(2,1)$ modes of the gravitational
    waveform in the inertial (top), coprecessing (middle), and coorbital
    (bottom) frames. In the inertial frame, the amplitude of the $(2,1)$ mode
    can be comparable to that of the $(2,2)$ mode. In the coprecessing frame,
    on the other hand, the $(2,2)$ mode always dominates. In addition, most
    effects of precession are removed by the rotation and the waveform in the
    coprecessing frame resembles that of a nonprecessing system. In the
    coorbital frame, finally, the waveform is further simplified and does not
    oscillate about zero. Mass ratio and initial spins used to produce this
    figure are indicated in the text within the figure.
}
\label{fig:precessing_frames}
\end{figure}

By contrast, for precessing systems the direction of $\bL$ varies due to
precession~\cite{Apostolatos:1994pre} and so there is not a fixed axis along
which the radiation is dominant. The standard practice is to choose $\hat{z}$
of the source frame along the direction of $\bL$ (or the total angular
momentum) at a reference time or frequency.

Heuristically, one can think of a precessing system as a nonprecessing system
with time-dependent frame rotations applied to it. In this non-inertial frame
the rotation causes mixing of power between modes of fixed $\ell$. For example,
the power of the $(2,\pm2)$ modes leaks into the $(2,\pm1)$ and $(2,0)$ modes.
This means that all $\ell=2$ modes can be dominant in
Eq.~(\ref{eq:spherical_harm}). While this rotating-frame picture ignores some
dynamical features such as nutation, it accounts for most of the effects of
precession in the waveform.

By the same logic, one could apply a time-dependent rotation to a precessing
system such that $\hat{z}$ always lies along $\hat{\bL}(t)$. In this
non-inertial frame, referred to as the coprecessing frame~\cite{Schmidt:2010it,
OShaughnessy:2011pmr, Boyle:2011gg}, the radiation is always strongest along
$\hat{z}$, and the ($\ell=2, m=\pm 2$) modes are dominant. In fact, since most
precessional effects are accounted for by the frame rotation, the waveform in
the coprecessing frame is qualitatively similar to that of a nonprecessing
system (cf.  Fig.~{\ref{fig:precessing_frames}}). This observation has been
exploited in the literature~\cite{Hannam:2013oca, Pan:2013rra, Khan:2018fmp,
    Blackman:2017dfb, Blackman:2017pcm} to simplify the modeling of precessing
    systems.  Here we proceed similarly, using the coprecessing frame described
    in Ref.~\cite{Boyle:2011gg} and denoting the strain in this frame as
    $\hlmCopr$.

\begin{figure*}[thb]
\includegraphics[width=0.9\textwidth]{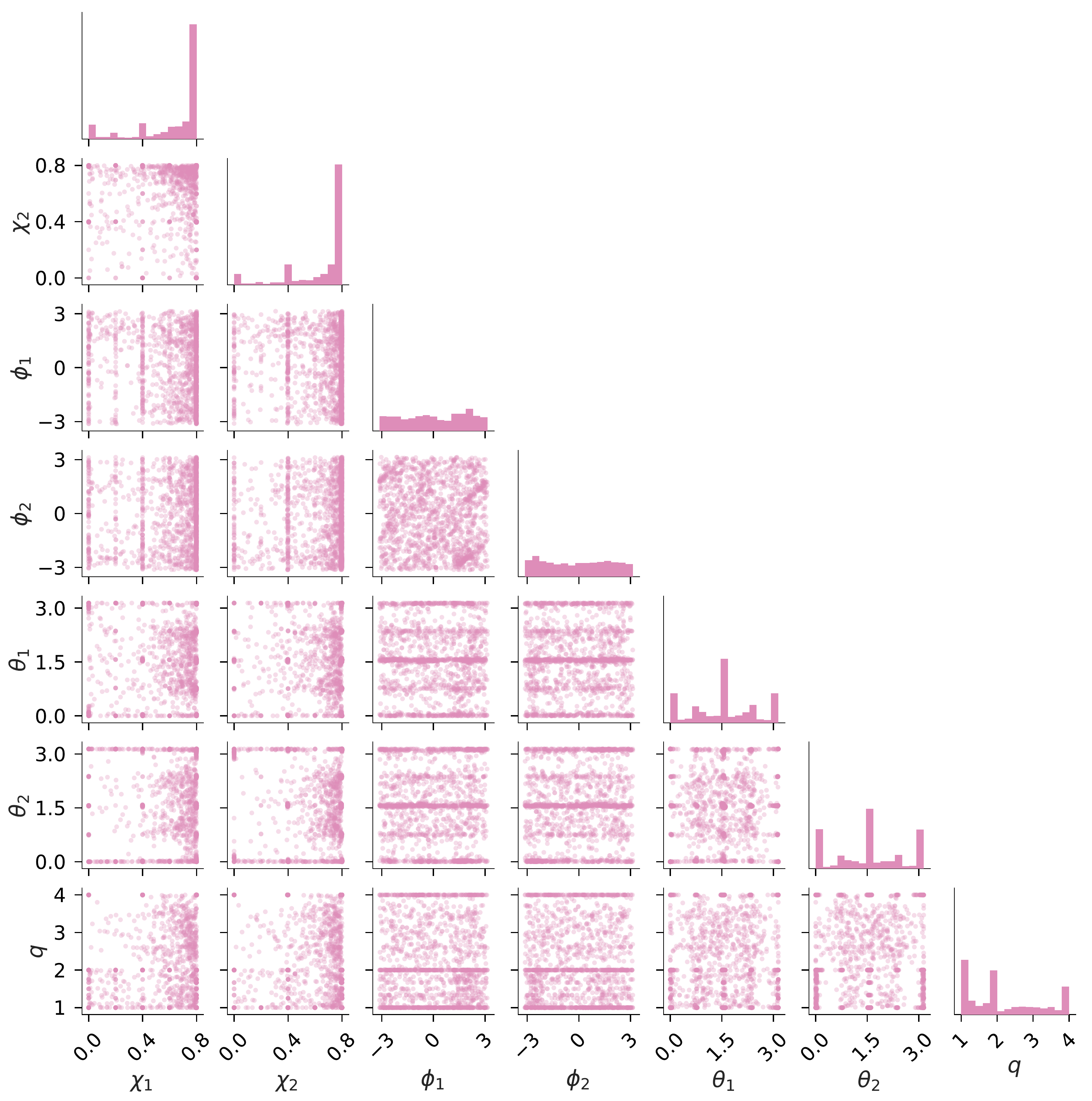}
\caption{Parameters of the 1528 NR simulations used in the construction of the
    surrogate models in this paper. We show the distribution of mass ratio $q$
    and the spin components in standard spherical polar coordinates ($\chi$,
    $\theta$, $\phi$) at $-4300M$ from the waveform amplitude peak. The
    index 1 (2) refers to the heavier (lighter) BH.
}
\label{fig:NR_params}
\end{figure*}

The waveform can be made even simpler, and therefore easier to model, by
applying an additional rotation about the $z-$axis of the coprecessing frame by
an amount equal to the instantaneous orbital phase:
\begin{equation}
    \hlmCoorb(t) = \hlmCopr(t)~e^{i m \phi(t)}.
\label{eq:coorb_hlm}
\end{equation}
Here we define the orbital phase,
\begin{equation}
    \phi(t) = \frac{\text{arg}[\hCopr_{2,-2}(t)]
    - \text{arg}[\hCopr_{2,2}(t)]}{4} \,,
    \label{eq:OrbitalPhase}
\end{equation}
using the coprecessing frame strain. The waveform $\hlmCoorb(t)$ corresponds to
a new frame, called the coorbital frame, in which the BHs are always on the
$x-$axis, with the heavier BH on the positive $x-$axis\footnote{Here the BH
    positions are defined from the waveform at future null infinity and do not
necessarily correspond to the (gauge-dependent) coordinate BH positions in the
NR simulation.}.  More importantly, the waveform in the coorbital frame is
nearly nonoscillatory, simplifying the modeling problem greatly.
Figure~\ref{fig:precessing_frames} shows an example of a waveform in the
inertial, coprecessing, and coorbital frames.

\section{NR simulations}
\label{sec:nr_simulations}

Our NR simulations are performed using the Spectral Einstein Code
(SpEC)~\cite{SpECwebsite, Pfeiffer:2002wt, Lovelace:2008tw, Lindblom:2005qh,
Szilagyi:2009qz, Scheel:2008rj} developed by the SXS~\cite{SXSWebsite}
collaboration.

\subsection{Parameter space coverage}

We use 890 precessing NR simulations used in the construction of the surrogate
models of Refs.~\cite{Blackman:2017pcm, Varma:2018aht}, which provide coverage
in the $q\leq2$ and $\chi_1,\chi_2 \leq 0.8$ regions of the parameter space. We
also make use of 64 aligned-spin simulations with $q\leq4$ and $\chi_1,\chi_2
\leq 0.8$ used in the construction of the surrogate model presented in
Ref.~\cite{Varma:2018mmi}. Finally, we performed 574 new simulations with $2 <
q \leq 4$, $\chi_1,\chi_2 \leq 0.8$ and generic spin directions---these
simulations are presented here for the first time. The parameters for the first
204 of these are chosen based on sparse grids as detailed in Appendix A of
Ref.~\cite{Blackman:2017pcm}. The remaining parameters are chosen as follows.
We randomly sample 1000 points uniformly in mass ratio, spin magnitude, and
spin direction on the sphere. We compute the distance between points a and b
using the metric
\begin{equation}
    ds^2 = \left( \frac{q^a - q^b}{\Delta q} \right)^2
    + \sum_{i\in\{1,2\}} \left( \frac{|\bchi^{a}_{i}
    - \bchi^{b}_{i}|}{\Delta \chi} \right)^2 ,
\end{equation}
where $\Delta q = 4 - 1 = 3$ and $\Delta \chi = 0.8$ are the ranges of these
parameters. These normalization factors are somewhat arbitrary, although any
choice of order unity should provide a reasonable criteria for point selection.
For each sampled parameter, we compute the minimum distance to all previously
chosen parameters. We then add the sampled parameter maximizing this minimum
distance to the set of chosen parameters. This is done iteratively for 370
additional parameters. The new simulations have identifiers SXS:BBH:1346-1350
and SXS:BBH:1514-2082, and are made publicly available through the SXS public
catalog~\cite{SXSCatalog}. The parameter space covered by the 890+64+574=1528
NR simulations used in this work is shown in Fig.~\ref{fig:NR_params}. Note
that not all of these are independent simulations: for 154 of these cases we
have $q=1$, with $\bchi_1 \neq \bchi_2$; for each of these cases we effectively
obtain an additional simulation by exchanging the labels of the two BHs.

The start time of these simulations varies between $4693 M$ and $5234 M$ before
the peak of the waveform amplitude, where $M = m_1 + m_2$ is the total
Christodoulou mass measured close to the beginning of the simulation at the
``relaxation time'' \cite{SXSCatalog2018}. The initial orbital parameters are
chosen through an iterative procedure~\cite{Buonanno:2010yk} such that the
orbits are quasicircular; the largest eccentricity for these simulations is
$9.8 \times 10^{-4}$, while the median value is $3.8 \times 10^{-4}$.

\subsection{Data extracted from simulations}

We make use of the following quantities extracted from the NR simulations: the
waveform modes $\hlm(t)$, the component spins $\bchi(t)$, the mass ratio $q$,
and the remnant mass $m_f$, spin $\bchi_f$, and kick velocity $\bv_f$.

The waveform is extracted at several extraction spheres at varying finite radii
from the origin and then extrapolated to future null
infinity~\cite{SXSCatalog2018, Boyle:2009vi}. Then the extrapolated waveforms
are corrected to account for the initial drift of the center of
mass~\cite{Boyle:2015nqa, scri}. The time steps during the simulations are
chosen nonuniformly using an adaptive time-stepper~\cite{SXSCatalog2018}. Using
cubic splines, we interpolate the real and imaginary parts of the waveform
modes to a uniform time step of $0.1 M$; this is dense enough to capture all
frequencies of interest, including near merger. The interpolated waveform at
future null infinity, scaled to unit mass and unit distance, is denoted as
$\hlm(t)$ in this paper.

The component spins $\bchi_{1,2}(t)$ and masses $m_{1,2}$ are evaluated on the
apparent horizons~\cite{Lovelace:2008tw} of the BHs. The masses at the
relaxation time~\cite{SXSCatalog2018} are used to define the mass ratio
$q=m_1/m_2$.  Unless otherwise specified, all masses in this paper are given in
units of the total mass $M=m_{1}+m_{2}$ at relaxation. The spins are
interpolated onto the same time array\footnote{The waveforms at future null
    infinity use a time coordinate $t$ that is different from the simulation
    time $\tilde{t}$ at which the spins are measured in the near
    zone~\cite{SXSCatalog2018}.  In this paper, we identify $t$ with
    $\tilde{t}$. While this identification is gauge-dependent, the spin
    directions are already gauge-dependent. We, however, note that the spin and
    orbital angular momentum vectors in the damped harmonic gauge used by SpEC
    agree quite well with the corresponding vectors in post-Newtonian (PN)
theory~\cite{Ossokine:2015vda}.} as used for the waveform, using cubic splines.

The remnant mass $m_{f}$ and spin $\bchi_{f}$ are determined from the common
apparent horizon long after ringdown, as detailed in
Ref.~\cite{SXSCatalog2018}. The remnant kick velocity is derived from
conservation of momentum, $\bm{v}_{f} = -\bm{P}^{\mathrm{rad}}/m_{f}$
\cite{Gerosa:2018qay}.  The radiated momentum flux $\bm{P}^{\mathrm{rad}}$ is
integrated~\cite{Ruiz:2007yx} from the strain $\hlm$.

\subsection{Post-processing the output of NR simulations}
\label{sec:post_processing}

After extracting the strain and spins from the simulations, we apply the
following post processing steps before building the surrogate models.

First, we shift the time arrays of all waveforms such that $t=0$ occurs at the
peak (see Ref.~\cite{Blackman:2017pcm} for how the peak is determined) of the
total waveform amplitude, defined as:
\begin{equation}
\label{eq:waveform_amplitude}
A(t)=\sqrt{\sum_{\ell m} |\hlm (t)|^2}.
\end{equation}

Then we rotate the waveform modes such that at a reference time $t_0=-4300M$,
the inertial frame coincides with the coorbital frame. This means that the
$\hat{z}$ direction of the inertial frame is along the principal eigenvector of
the angular momentum operator~\cite{Boyle:2011gg} at the reference time. In
addition, the $\hat{x}$ direction of the inertial frame is along the line of
separation from the lighter BH to the heavier BH (in other words, the orbital
phase is zero). The spin vectors $\bchi_{1,2}(t)$ are also transformed into the
same inertial frame.

We then truncate the waveform and spin time series by dropping all times
$t<-4300M$ to exclude the initial transients known as ``junk radiation''.
After the truncation, the reference time $t=-4300M$ is also the start time of
the data.

For $t>-100M$, the spin measurements from the apparent horizons start to become
unreliable as the horizons become highly distorted. Following
Ref.~\cite{Blackman:2017pcm}, starting at $\tmHundred$, we extend the spins to
later times using PN spin evolution equations. This evolution is done even past
the merger stage, into the ringdown. We stress that the extended spins are
unphysical but are a useful parametrization to construct fits at late times.

Finally we apply a smoothing filter (see Eq.~(6) of
Ref.~\cite{Blackman:2017pcm}) on the spin time series to remove fast
oscillations taking place on the orbital timescale. This smoothing helps
improve the numerical stability of the ordinary differential equation (ODE)
integrations described in Sec.~\ref{sec:dynamics_sur}. Note that we use the
filtered spins for the waveform surrogate (Sec.~\ref{sec:waveform_sur}) but not
for the remnant surrogate (Sec.~\ref{sec:remnant_sur}), for which we just use
the unfiltered spins since there are no ODE integrations involved.

\section{Waveform surrogate}
\label{sec:waveform_sur}

To construct the waveform surrogate, we closely follow the model of
Ref.~\cite{Blackman:2017pcm}, with some modifications to adapt it to higher
mass ratios. We refer to the new waveform model as NRSur7dq4.

\begin{figure}[thb]
\includegraphics[width=0.5\textwidth]{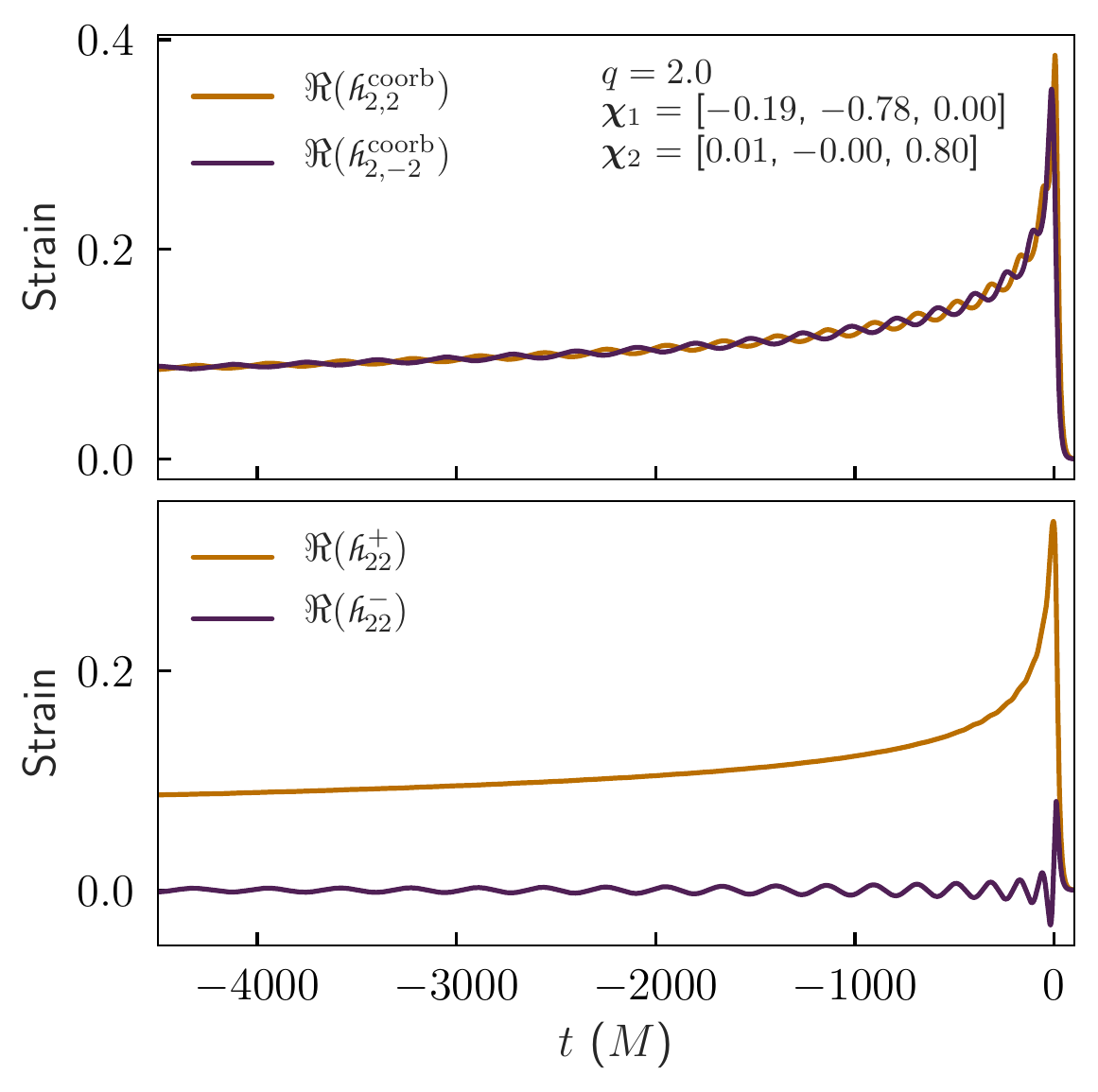}
\caption{The top panel shows the real part of the $(2,2)$ and $(2,-2)$ modes of
    the waveform in the coorbital frame. Notice that the orbital time scale
    oscillations of these two modes have opposite signs. The bottom panel shows
    the real parts of $\h^{+}_{2,2}$ and $\h^{-}_{2,2}$ (cf.
    Eq.~\ref{eq:plus_minus_combinations}), we take advantage of the above fact
    to move most of the oscillations from the larger to the smaller data piece.
}
\label{fig:plus_minus_combinations}
\end{figure}

\subsection{Coorbital frame surrogate}
\label{sec:coorb_sur}

We find that the surrogate accuracy improves when working with slowly varying
functions, rather than oscillatory ones. Therefore, we first decompose the
strain into several ``data pieces'', each of which is a slowly varying function
of time, and build a surrogate for each of them.  At evaluation time, we
combine the various data pieces to reconstruct the inertial frame strain. To
reduce the cost of these transformations, we first downsample the inertial
frame strain onto a set of 2000 time values $\tCoorb_i$ that are approximately
uniformly spaced in the orbital phase (using the method described in App.~B of
Ref.~\cite{Blackman:2017pcm}).

As described in Sec.~\ref{sec:prelims}, the waveform is simpler in the
coorbital frame. A further simplification is possible by considering
combinations of $m>0$ and $m<0$ counterparts of a fixed $\ell$ mode:
\begin{equation}
    \hlm^{\pm} = \frac{\hCoorb_{\ell, m} \pm \hCoorb_{\ell, -m} \,^*}{2}.
\label{eq:plus_minus_combinations}
\end{equation}
Figure~\ref{fig:plus_minus_combinations} shows an example of the simplification
obtained with this combination. For all $m\neq0$ modes we model the real and
imaginary parts of $\hlm^{\pm}$. For $m=0$ modes, we directly model the real
and imaginary parts of the coorbital frame strain $\hCoorb_{\ell, m}$.  We
construct an independent surrogate model for each of these data pieces and
refer to the combination of these models as the coorbital frame surrogate.

As described in Ref.~\cite{Blackman:2017pcm}, for each waveform data piece, we
construct a linear basis using singular value decomposition with an RMS
tolerance of $3\times10^{-4}$. We then construct an empirical time interpolant
with the same number of empirical time nodes as basis functions for that data
piece~\cite{Field:2013cfa,Maday:2009,chaturantabut2010nonlinear}.  The
empirical time nodes are chosen as a subset of the 2000 coorbital time values
($\tCoorb_i$) and are specific to each data piece.  Finally, for each empirical
time node, we construct a parametric fit for the waveform data piece. The fits
are parametrized as functions of the mass ratio and the spins in the coorbital
frame at that time. We describe our fitting procedure in
Sec.~\ref{sec:parametric_fits_wavesur}. At evaluation time, the coorbital frame
spins at any time are obtained using the dynamics surrogate described in
Sec.~\ref{sec:dynamics_sur}.

\subsection{Dynamics surrogate}
\label{sec:dynamics_sur}

The surrogate described in Sec.~\ref{sec:coorb_sur} only models the strain in
the coorbital frame. We also need to model the following quantities:

\begin{enumerate}
\item The orbital phase in the coprecessing frame, which is required to
    transform the strain from the coorbital frame to the coprecessing frame
    [cf. Eq.~(\ref{eq:coorb_hlm})];
\item The quaternions describing the coprecessing frame, which are required to
    transform the strain from the coprecessing frame to the inertial frame;
\item The spins as a function of time, which are used in the
    evaluation of the parametric fits described in
    Sec.~\ref{sec:parametric_fits_wavesur}.
\end{enumerate}

We refer to the model for these quantities as the dynamics surrogate. Using the
fitting method of Sec.~\ref{sec:parametric_fits_wavesur}, we first construct
parametric fits for $\omega(t)$, $\OmegaCoorb_{x,y}(t)$, and
$\bchidotCoorb_{1,2}(t)$ at selected time nodes referred to as the dynamical
time nodes $\tDyn_i$.  Here $\bchidotCoorb_{1,2}(t)$ are the time derivatives
of the coprecessing frame spins transformed to the coorbital frame, $\omega(t)$
is ${d\phi}/{dt}$ (cf. Eq.~(\ref{eq:OrbitalPhase})), and $\OmegaCoorb_{x,y}(t)$
is the angular velocity of the coprecessing frame, transformed to the coorbital
frame.  These quantities are described in more detail in Sec.~III of
Ref.~\cite{Blackman:2017pcm}.  Note that $\OmegaCoorb_{z}(t)\sim0$.  For the
dynamical time nodes $\tDyn_i$ we chose 238 time values such that there are
approximately $10$ nodes per orbit (see App. B of Ref.~\cite{Blackman:2017pcm}
for details).

We use a fourth-order Adams-Bashforth scheme to integrate $\omega(\tDyn_i)$,
$\OmegaCoorb_{x,y}(\tDyn_i)$, and $\bchidotCoorb_{1,2}(\tDyn_i)$ over the set
of dynamical time nodes $\tDyn_i$ providing the time evolution of the orbital
phase $\phi(\tDyn_i)$, the coprecessing frame quaternions $\Q(\tDyn_i)$, and
the component spins in the coorbital frame $\bchiCoorb_{1,2}(\tDyn_i)$. This
involves solving  a coupled ODE as described in Sec.~V of
Ref.~\cite{Blackman:2017pcm}. At each step of the ODE integration, the
coorbital frame spins at the current node $\tDyn_i$ are first obtained. These
are then used to evaluate the parametric fits for the derivative quantities
mentioned above.  Note that the spins used in the dynamics surrogate are the
filtered spins mentioned in Sec.~\ref{sec:post_processing}; this improves the
accuracy of the ODE integration by making the spin time derivatives easier to
model.

\subsection{Parametric fits}
\label{sec:parametric_fits_wavesur}

For the coorbital frame surrogate of Sec.~\ref{sec:coorb_sur}, we need to
construct parametric fits at various empirical time nodes for the different
data pieces. Similarly, for the dynamics surrogate of
Sec.~\ref{sec:dynamics_sur}, we need to construct fits for various time
derivatives at the dynamical time nodes $\tDyn_i$. We use the same procedure
for each of these fits. Let us refer to the data to be fitted as $y(\bLambda)$,
where $\bLambda$ is a seven-dimensional set of parameters.

For each of these fits, the seven parameters $\bLambda$  must contain
information on mass ratio $q$ and coorbital frame spins $\bchiCoorb_{1,2}(t_i)$
at the time corresponding to the fit.  Following Ref.~\cite{Varma:2018aht}, we
parametrize the fits using
\begin{equation}
\bLambda=[\log(q), \chiCoorb_{1x},
\chiCoorb_{1y}, \chihatCoorb, \chiCoorb_{2x}, \chiCoorb_{2y}, \chiCoorb_{a}]\,,
\label{bLambda}
\end{equation}
where $\chihatCoorb$ is the spin parameter entering the GW phase at leading
order \cite{Khan:2015jqa, Ajith:2011ec, CutlerFlanagan1994, Poisson:1995ef} in
the PN expansion
\begin{gather}
\chihatCoorb = \frac{\chieffCoorb - 38\eta(\chiCoorb_{1z}
     + \chiCoorb_{2z})/113} {1-{76\eta}/{113}} \, , \\
\chieffCoorb = \frac{q~\chiCoorb_{1z} + \chiCoorb_{2z}}{1+q} , \\
\eta = \frac{q}{(1+q)^2} \, ,
\end{gather}
and $\chiCoorb_a$ is the ``anti-symmetric spin'',
\begin{equation}
    \chiCoorb_a = \tfrac{1}{2}(\chiCoorb_{1z} - \chiCoorb_{2z})\,.
\label{eq:anti_symm_spin}
\end{equation}
We empirically found this parameterization to perform more accurately than the
more intuitive choice $\bLambda_\mathrm{ref56}=[q, \chiCoorb_{1x},
\chiCoorb_{1y}, \chiCoorb_{1z}, \chiCoorb_{2x}, \chiCoorb_{2y},
\chiCoorb_{2z}]$ used in Ref.~\cite{Blackman:2017pcm}.

Fits are constructed using the forward-stepwise greedy fitting method described
in App. A of Ref.~\cite{Blackman:2017dfb}. We choose the basis functions to be
a tensor product of 1D monomials in the components of $\bLambda$. The
components of $\bLambda$ are first affine mapped to the interval $[-1,1]$
before constructing the tensor product. We consider up to cubic powers in
$\log(q)$ and up to quadratic powers in the spin parameters. We find that going
to higher powers does not significantly improve the fit accuracy within the
training region, but the mass ratio extrapolation errors estimated in
App.~\ref{sec:extrap} become much larger.

It is always possible to improve the accuracy of a fit by adding more basis
functions. However, this can lead to over-fitting when the data contain some
noise. Our source of noise is mostly due to NR truncation error, but also
systematic errors such as waveform extrapolation and residual eccentricity. In
order to safeguard against over-fitting, we perform 10 trial fits, leaving a
random $10\%$ of the dataset out as validation points in each trial, to
determine the set of basis functions used in constructing the final fit. We
allow a maximum of 100 basis functions for each fit. See App.~A of
Ref.~\cite{Blackman:2017dfb} for more details.

\subsection{Surrogate evaluation}

To evaluate the surrogate, we begin with a user-specified mass ratio $q$ and
spins $\bchiCoorb_{1,2}$ at the initial time $t=-4300M$. Note that at this
time, the inertial frame coincides with the coorbital frame.  These values are
used to initialize the dynamics surrogate described in
Sec.~\ref{sec:dynamics_sur}, which predicts the coprecessing frame quaternions
$\Q(\tDyn_i)$, the orbital phase $\phi(\tDyn_i)$ in the coprecessing frame, and
the coorbital frame spins $\bchiCoorb_{1,2}(\tDyn_i)$ at the dynamic time nodes
$\tDyn_i$. We then use cubic splines to interpolate these quantities on to the
time array for the coorbital frame surrogate $\tCoorb_i$, giving us
$\Q(\tCoorb_i)$, $\phi(\tCoorb_i)$, and $\bchiCoorb_{1,2}(\tCoorb_i)$.

The coorbital frame surrogate described in Sec.~\ref{sec:coorb_sur} is used to
predict the strain in the coorbital frame.  This involves evaluating the fits
at the empirical time nodes for this surrogate using
$\bchiCoorb_{1,2}(\tCoorb_i)$ and $q$.  Then, the orbital phase
$\phi(\tCoorb_i)$ is used to transform the strain from the coorbital frame to
the coprecessing frame (cf. Eq.~\ref{eq:coorb_hlm}). Finally, the coprecessing
frame quaternions $\Q(\tCoorb_i)$ are used to transform the strain from the
coprecessing frame to the inertial frame (this involves Wigner matrices, see
App. A of Ref.~\cite{Boyle:2011gg}). This gives us $\hlm(\tCoorb_i)$, which is
interpolated onto any required time array $t$ using cubic splines to get
$\hlm(t)$.

\section{Remnant surrogate}
\label{sec:remnant_sur}

To construct the remnant properties surrogate, we closely follow the model of
Ref.~\cite{Varma:2018aht}. We refer to the new model presented here as
\RemnantModel.

We model the remnant mass $m_f$, spin $\bchi_f$, and kick velocity $\bv_f$.
Before constructing the fits, $\bchi_f$ and $\bv_f$ are transformed into the
coorbital frame at $\tmHundred$. We model each component of the vectors
independently. The fits are parametrized by the same $\bLambda$ of
Eq.~\eqref{bLambda}, but using the component spins at $\tmHundred$.  Unlike the
waveform surrogate case, we do not filter out orbital-timescale oscillations.
The filtered spins were found to be necessary for the accuracy of the time
integration in Sec.~\ref{sec:dynamics_sur}, which is not necessary here because
the remnant properties can evaluated from the BBH parameters at a single time
$\tmHundred$.

All fits are performed using Gaussian Process Regression (GPR), as described in
the supplementary materials of Ref.~\cite{Varma:2018aht}. We find that GPR
fitting is, in most cases, more accurate but also significantly more expensive
than the polynomial fitting method described in
Sec.~\ref{sec:parametric_fits_wavesur}. GPR becomes impractical to use for the
waveform surrogate as there are hundreds of fits that need to be evaluated to
generate the waveform. For the remnant fits, however, the additional cost of
GPR is acceptable because one is only fitting 7 quantities ($m,\bchi_f,\bv_f$).
In addition, GPR naturally provides error estimates which can be useful in data
analysis applications.  The efficacy of the GPR error estimate in reproducing
the underlying error of the surrogate models was investigated thoroughly in the
supplementary materials of Ref.~\cite{Varma:2018aht}.

Although \RemnantModel is parameterized internally by input spins specified in
the coorbital frame at $\tmHundred$, we allow the user to specify input spins
at earlier times, and in the inertial frame; this case is handled by two
additional levels of spin evolution.  Given the inertial-frame input spins at
an initial orbital frequency $f_0$, we first evolve the spins using a
post-Newtonian (PN) approximant --- 3.5PN SpinTaylorT4 \cite{Buonanno:2002fy,
Boyle:2007ft, Ossokine:2015vda} --- until we reach the domain of validity of
the more accurate NRSur7dq4 ($t=-4300M$ from the peak).  We then use the
dynamics surrogate of NRSur7dq4 to evolve the spins until $\tmHundred$. These
spins are then transformed to the coorbital frame and used to evaluate the
remnant fits. Thus, spins can be specified at any given orbital frequency and
are evolved consistently before estimating the final BH properties. Note that
NRSur7dq4 uses the filtered spins, while \RemnantModel expects unfiltered spins
at $\tmHundred$, but we find that the errors introduced by this discrepancy are
negligible compared to the errors due to PN spin evolution.

\section{Results}
\label{sec:results}

\begin{figure*}[t!]
    \centering
    \begin{subfigure}[t]{0.5\textwidth}
        \centering
        \includegraphics[width=0.9\textwidth]{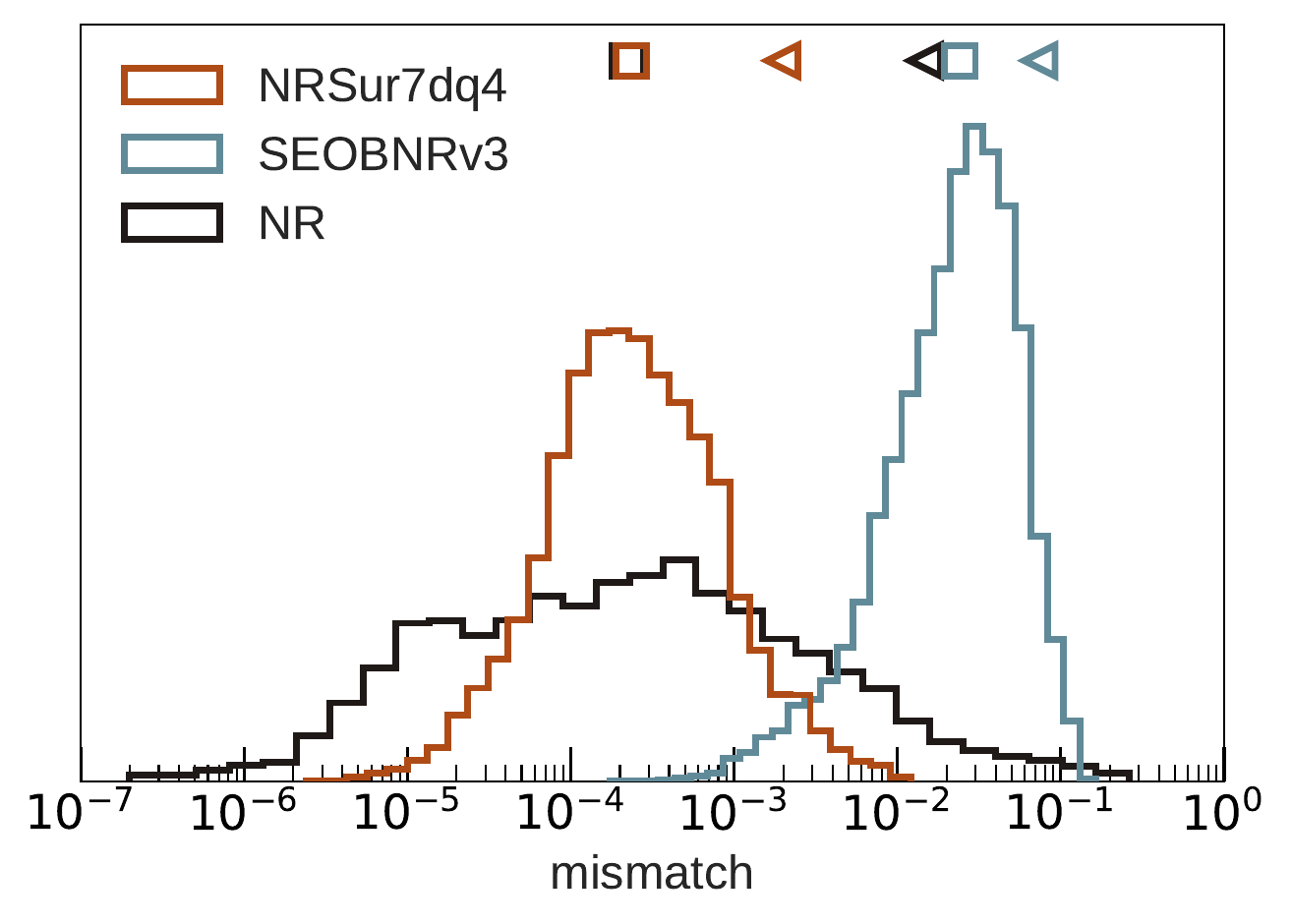}
        \caption{Flat noise mismatches}
        \label{fig:waveform_errors_flat}
    \end{subfigure}%
    ~ 
    \begin{subfigure}[t]{0.5\textwidth}
        \centering
        \includegraphics[width=0.89\textwidth]{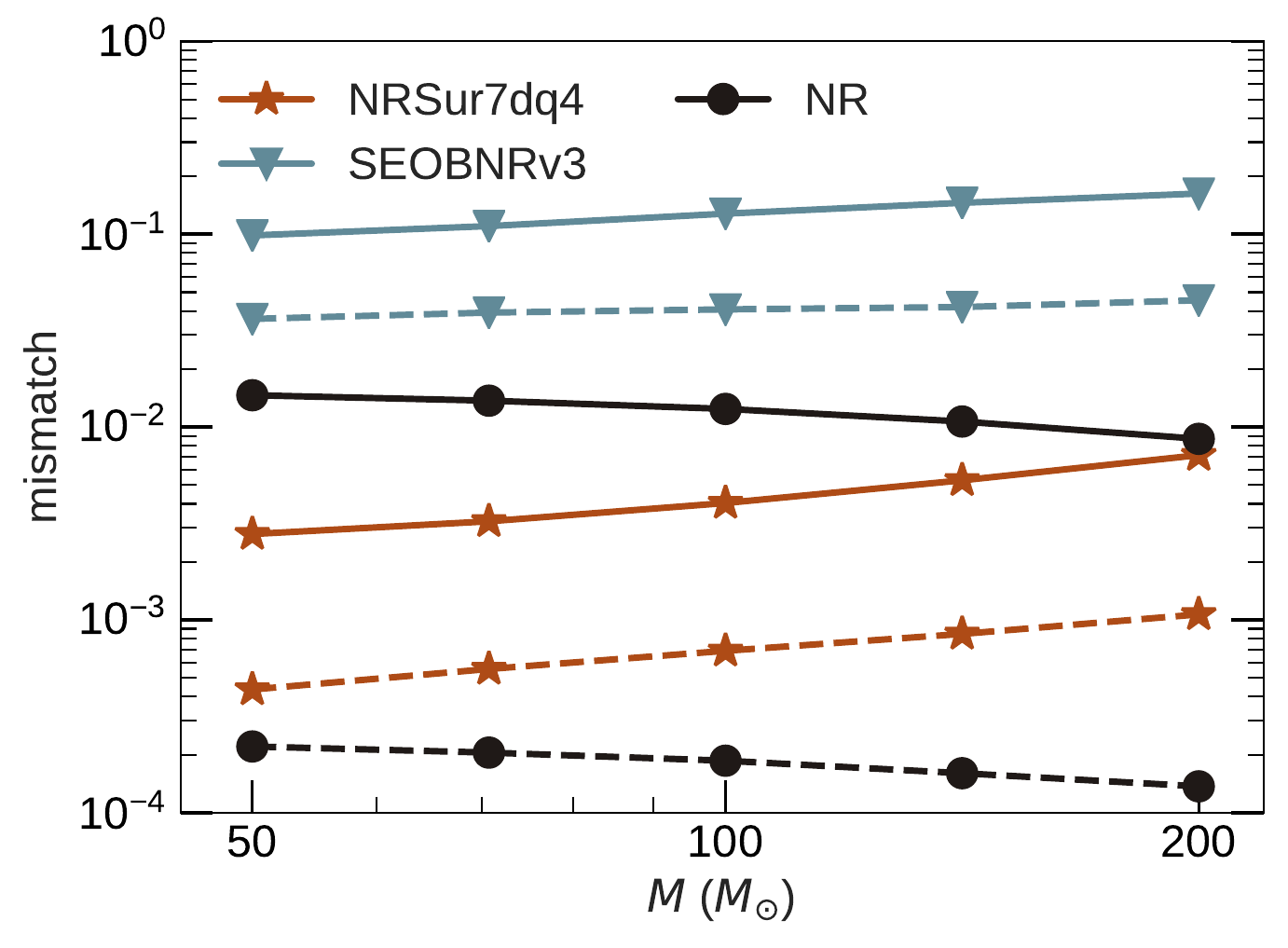}
        \caption{LIGO noise mismatches}
        \label{fig:waveform_errors_ligo}
    \end{subfigure}
    \caption{Mismatches for NRSur7dq4 and \eobmodel models, when compared
        against precessing NR simulations using all $\ell\leq5$ modes with mass
        ratios $q\leq4$, and spin magnitudes $\chi_{1},\chi_{2}\leq0.8$. The
        NRSur7dq4 errors shown are out-of-sample errors. Also shown are the NR
        resolution errors. Mismatches are computed at several sky locations
        using all available modes for each model: $\ell\leq4$ for NRSur7dq4,
        and $\ell=2$ for \eobmodel. The NR error is computed using all
        $\ell\leq5$ modes from the two highest available resolutions.
        \textit{Left panel}: Mismatches computed using a flat noise curve.  The
        square (triangle) markers at the top indicate the median ($95$th
        percentile) values.  \textit{Right panel}: Mismatches computed using
        the Advanced LIGO design sensitivity noise curve, as a function of
    total mass.  The dashed (solid) lines indicate the median ($95$th
percentile) values over different NR simulations and points in the sky.}
    \label{fig:waveform_errors}
$\,$\\ $\,$\\
\capstart
\includegraphics[width=0.9\textwidth]{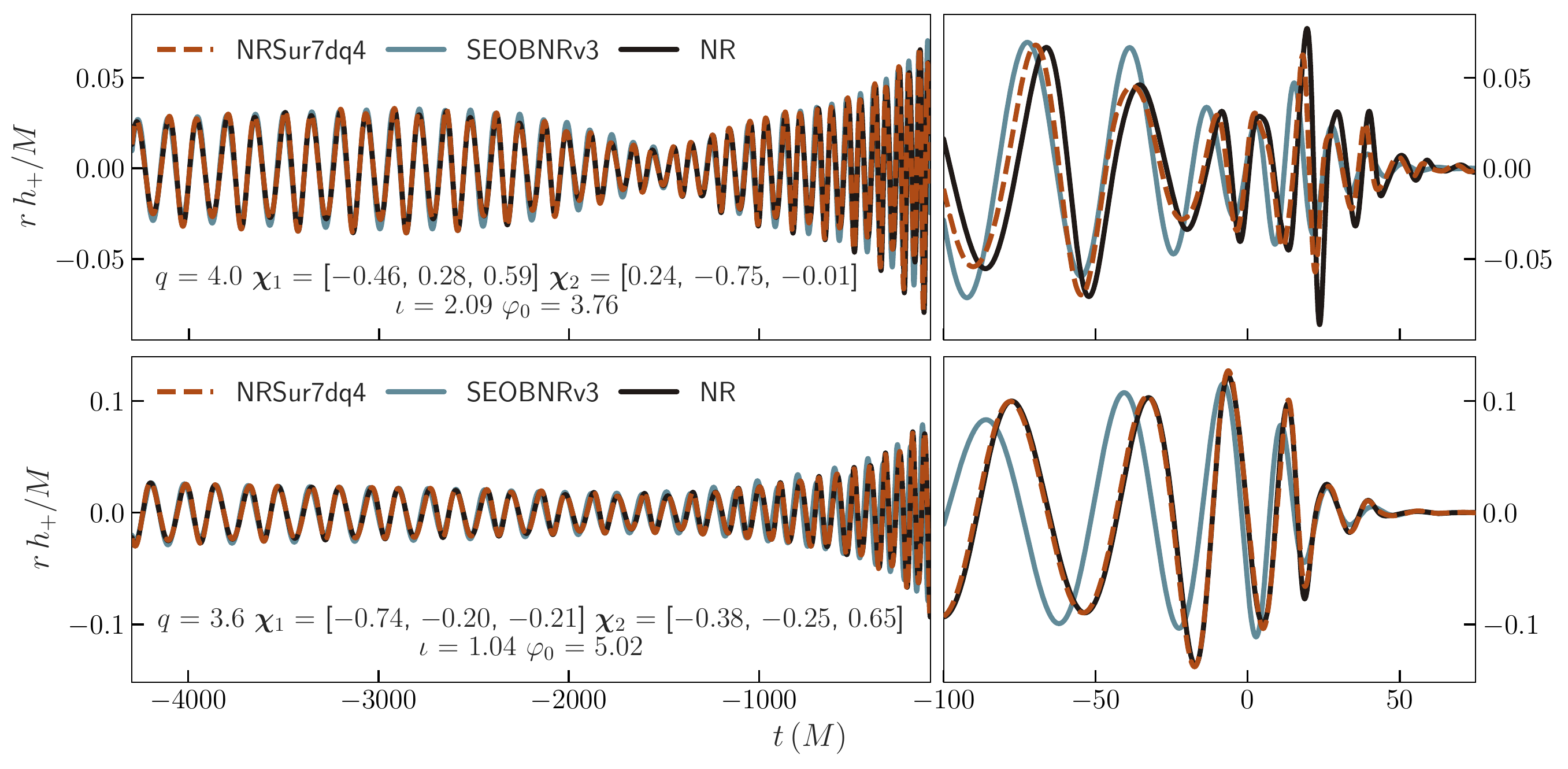}
\caption{The plus polarization of the waveforms for the cases that result in
    the largest mismatch for NRSur7dq4 (top) and \eobmodel (bottom) in
    Fig.~\ref{fig:waveform_errors_flat}. We also show the corresponding NR
    waveforms. Each waveform is projected using all available modes for that
    model, along the direction that results in the largest mismatch for
    NRSur7dq4 (\eobmodel) in the top (bottom) panel. Note that NRSur7dq4 is
    evaluated using trial surrogates that are not trained using these cases.
    The binary parameters and the direction in the source frame are indicated
    in the figure text. All waveforms are time shifted such that the peak of
    the total amplitude occurs at t = 0 [using all available modes, according
    to Eq.~(\ref{eq:waveform_amplitude})]. The waveform modes are then rotated
    to have their orbital angular momentum aligned with the $z$-axis, and such
    that the orbital phase is equal to zero at $t=-4300M$.
}
\label{fig:worst_case}
\end{figure*}

We evaluate the accuracy of our new surrogate models by comparing against the
waveform and remnant properties from the NR simulations used in this work.  For
this, we perform a 20-fold cross-validation study to compute ``out-of-sample''
errors as follows. We first randomly divide the 1528 training simulations into
20 groups of $\roughly 76$ simulations each.  For each group, we build a trial
surrogate using the $\roughly 1452$ remaining training simulations and test
against these $\roughly 76$ validation ones, which may include points on the
boundary of the training set.

\subsection{Waveform surrogate errors}
\label{sec:waveform_errs}

To estimate the difference between two waveforms, $\h_1$ and $\h_2$, we use the
mismatch
\begin{gather}
\mathcal{MM} = 1 - \frac{\left<\h_1, \h_2\right>}{\sqrt{\left<\h_1, \h_1\right>
    \left<\h_2, \h_2\right>}}, \\
\left<\h_1, \h_2\right> = 4 \mathrm{Re}
    \int_{f_{\mathrm{min}}}^{f_{\mathrm{max}}}
    \frac{\tilde{\h}_1 (f) \tilde{\h}_2^* (f) }{S_n (f)} df,
\label{eq:mismatch}
\end{gather}
where $\tilde{\h}(f)$ indicates the Fourier transform of the complex strain
$\h(t)$, $^*$ indicates a complex conjugation, $\mathrm{Re}$ indicates the real
part, and $S_n(f)$ is the one-sided power spectral density of a GW detector. We
taper the time domain waveform using a Planck window~\cite{McKechan:2010kp},
and then zero-pad to the nearest power of two.  We further zero-pad the
waveform to increase the length by a factor of eight before performing the
Fourier transform. The tapering at the start of the waveform is done over $1.5$
cycles of the $(2,2)$ mode. The tapering at the end is done over the last
$30M$. Note that our model contains times up to $100M$ after the peak of the
waveform amplitude, and the signal has essentially died down by the last $30M$.
We take $f_{\mathrm{min}}$ to be twice the waveform angular velocity (as
defined by Ref.~\cite{Boyle:2013nka}) at the end of the initial tapering
window, and $f_{\mathrm{max}}$ is chosen to be 4 times the waveform angular
velocity at $t=0$; the extra factor of 4 is chosen to resolve up to $m=4$
spherical-harmonic modes, with an extra margin of a factor of 2. We compute
mismatches with a flat noise curve ($S_n=1$) as well as with the Advanced-LIGO
design sensitivity noise curve~\cite{aLIGODesignNoiseCurve}. Mismatches are
computed following the procedure described in Appendix D of
Ref.~\cite{Blackman:2017dfb}. In particular, we optimize over shifts in time,
polarization angle, and initial orbital phase.  Both plus and cross
polarizations are treated on an equal footing by using a two-detector setup
where one detector sees only the plus and the other only the cross
polarization. We compute the mismatches at 37 points uniformly distributed on
the sky in the source frame, and we use all available modes of a given waveform
model.

Figure~\ref{fig:waveform_errors} summarizes the out-of-sample mismatches for
NRSur7dq4 against the NR waveforms. In Fig.~\ref{fig:waveform_errors_flat} we
show mismatches computed using a flat noise curve.  We compare this with the
truncation error in the NR waveforms themselves, estimated by computing the
mismatch between the two highest available resolutions of each NR simulation.
The errors in the surrogate model are well within the estimated truncation
errors of the NR simulations. In addition, we also show the errors for the
waveform model \eobmodel~\cite{Pan:2013rra, Babak:2016tgq}, which also includes
spin precession effects~\footnote{Note that \eobmodel spins are specified at a
    reference frequency, rather than a time before merger. We choose the
reference frequency such that the waveform begins at $t=-4300M$ before the
waveform amplitude peak (as defined in Eq.~\ref{eq:waveform_amplitude}).}.  The
surrogate errors are at least an order of magnitude lower than those of
\eobmodel.

Apart from \eobmodel, another model commonly used in data analysis applications
is IMRPhemomPv2~\cite{Hannam:2013oca}. IMRPhemomPv2 was shown to be comparable
in accuracy to SEOBNRv3 in Ref.~\cite{Blackman:2017pcm}, at least in order of
magnitude. Therefore, for simplicity, we do not show comparisons of
IMRPhemomPv2 to NR here. Note that updated versions of both SEOBNRv3 (based on
Ref.~\cite{Cotesta:2018fcv}) and IMRPhemomPv2 (see Ref.~\cite{Khan:2018fmp})
are under development, but are not currently available publicly. We note that
these models are calibrated only against aligned-spin NR simulations, using a
much smaller set of simulations than our model. Both these factors contribute
to the accuracy of these models. On the other hand, these models are expected
to be valid for larger mass ratios and spin magnitudes than our model, although
their accuracy in that region is unknown due to lack of sufficient number of
simulations.

We note that the NR truncation mismatch distribution in
Fig.~\ref{fig:waveform_errors_flat} has a tail extending to $\mathcal{MM}\sim
0.1$.  We find that these cases occur when the spins of the two highest
resolutions of the simulation are inconsistent with each other because of
unresolved effects during junk-radiation emission, meaning that the two
resolutions represent different physical systems. This means that comparing the
resolutions for these cases gives us an error estimate that is too conservative
and does not reflect the actual truncation error of the simulations. We expect
the actual truncation error to be closer to the errors reproduced by the
surrogate model (which is trained on the high resolution data set) in
Fig.~\ref{fig:waveform_errors_flat}.  Evidence for these claims is provided in
App.~\ref{sec:PBandJ}.

\begin{figure}[hbt]
\includegraphics[width=0.45\textwidth]{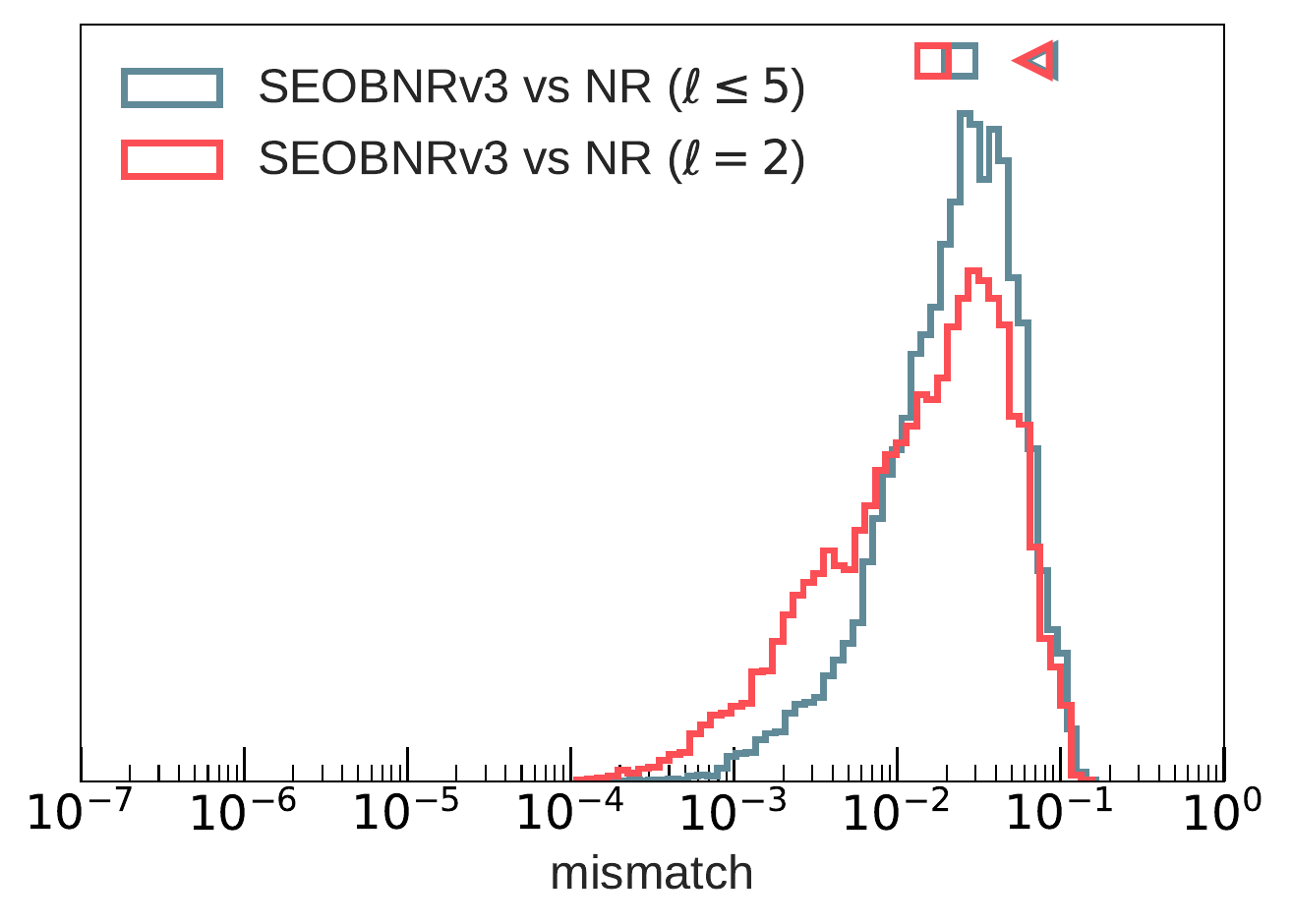}
\caption{Same as Fig.~\ref{fig:waveform_errors_flat} but using only $\ell=2$
    modes for NR when compared to \eobmodel. The blue histogram from
    Fig.~\ref{fig:waveform_errors_flat}, where \eobmodel is compared to NR with
    all $\ell\leq5$ modes, is reproduced here for comparison. The square
    (triangle) markers at the top indicate the median ($95$th percentile)
    values.
}
\label{fig:waveform_errors_ell2}
\end{figure}

\begin{figure*}[thb]
\includegraphics[width=0.9\textwidth]{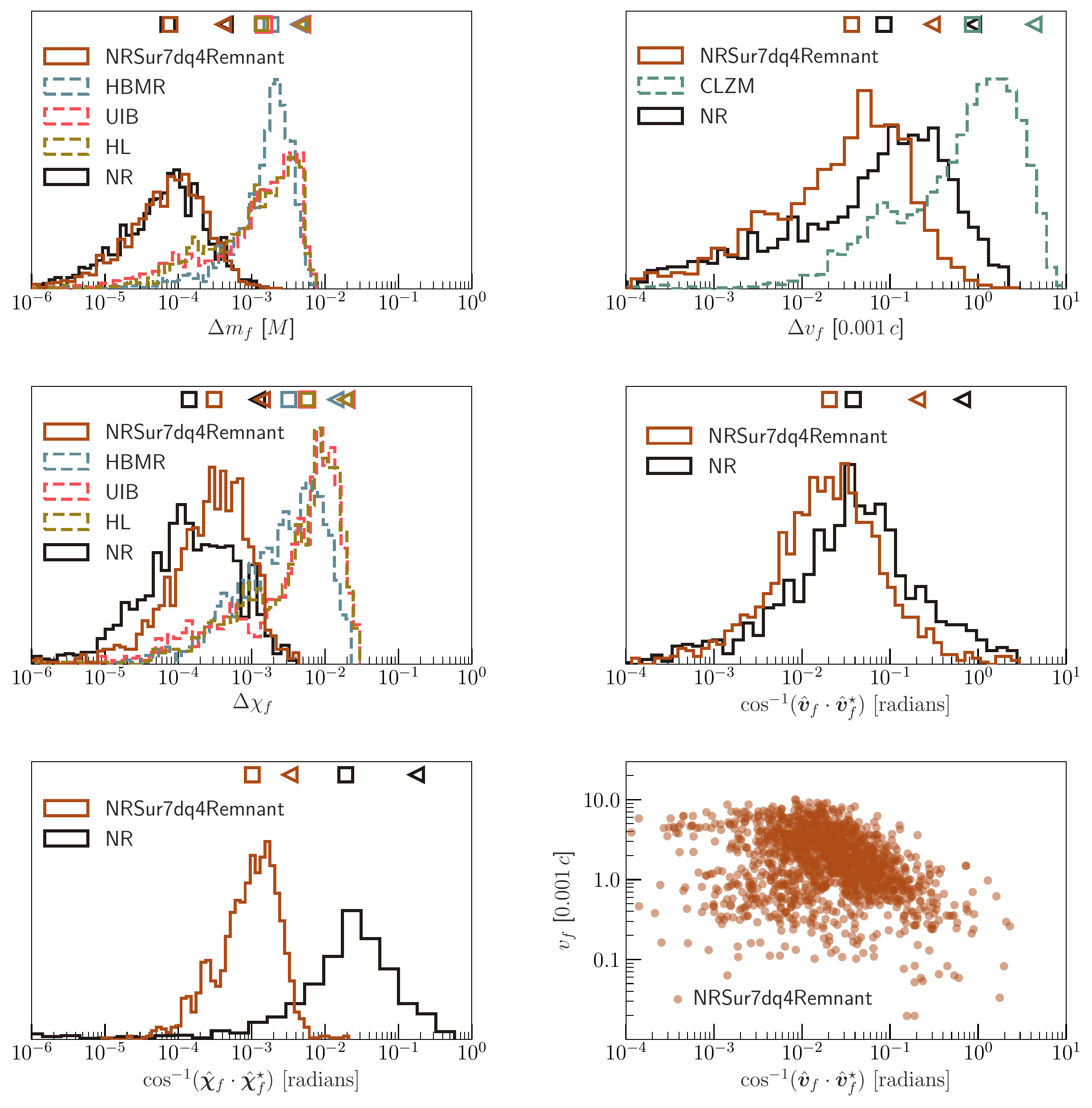}
\caption{Error histograms for \RemnantModel for the remnant mass, spin
    magnitude, spin direction, kick magnitude, and kick direction for
    precessing BBH with mass ratios $q\leq4$ and spin magnitudes
    $\chi_{1},\chi_{2}\leq0.8$.  The direction error is the angle between the
    predicted vector and a fiducial vector, taken to be the high-resolution NR
    case and indicated by $^{\star}$. Square (triangle) markers indicate median
    ($95$th percentile) values.  Also shown are the NR resolution errors and
    errors for different existing fitting formulae. In the bottom-right panel
    we show the joint distribution of kick magnitude and kick-direction error.
}
\label{fig:remnant_errors}
\end{figure*}

Fig.~\ref{fig:waveform_errors_ligo} shows mismatches computed using the
Advanced LIGO design sensitivity noise curve~\cite{aLIGODesignNoiseCurve}. In
this case,  results depend on the total mass $M$ of the system. Consequently,
we show the median and 95th percentile values at different $M$, rather than
full histograms. Once again, the surrogate errors are comparable to those of
the NR simulations, and are at least an order of magnitude lower than that of
\eobmodel. Over the mass range $50-200 M_{\odot}$, mismatches for NRSur7dq4 are
always $\lesssim8\times10^{-3}$ at the $95$ percentile level.

Fig.~\ref{fig:worst_case} shows a comparison of waveforms computed via
NRSur7dq4, \eobmodel, and NR for the cases that lead to the largest error for
NRSur7dq4 and \eobmodel in Fig.~\ref{fig:waveform_errors_flat}.  The surrogate
shows reasonable agreement with NR, even for its worst case, while \eobmodel
shows a noticeably larger deviation in both cases.

In Figs.~\ref{fig:waveform_errors} and \ref{fig:worst_case} we use all
available modes for NRSur7dq4 and \eobmodel. NRSur7dq4 models all modes
$\ell\leq4$, while \eobmodel models only the $\ell=2$ modes. For the NR
waveforms in Figs.~\ref{fig:waveform_errors} and \ref{fig:worst_case}, we
include all modes $\ell\leq5$ to account for the error due to neglecting
$\ell>4$ modes in NRSur7dq4. To better understand what fraction of the
\eobmodel error comes from neglecting modes with $\ell>2$, we repeat the
calculations leading to the \eobmodel histogram in
Fig.~\ref{fig:waveform_errors_flat} in Fig.~\ref{fig:waveform_errors_ell2},
while restricting all waveforms to $\ell=2$. While there is a noticeable move
towards lower mismatches when restricted to $\ell=2$, the median and $95$th
percentile values change only marginally, suggesting that the main error source
for \eobmodel are the $\ell=2$ modes themselves.

\subsection{Remnant surrogate errors}
\label{sec:remnant_errs}

We evaluate the accuracy of the remnant surrogate \RemnantModel by comparing
against the NR simulations through a cross-validation study as in
Sec.~\ref{sec:waveform_errs}. Out-of-sample errors for the remnant properties
predicted by \RemnantModel are shown in Fig.~\ref{fig:remnant_errors}. $95$th
percentile errors are $\roughly 5 \into 10^{-4} M$ for mass, $\roughly 2 \into
10^{-3}$ for spin magnitude, $\roughly 4 \into 10^{-3}$ radians for spin
direction, $\roughly 4 \into 10^{-4}\,c$ for kick magnitude, and $\roughly 0.2$
radians for kick direction.  Our errors are at the same level as the NR
resolution error, estimated by comparing the two highest NR resolutions.  The
largest errors in the kick direction can be of order $\roughly 1$ radian. The
bottom-right panel of Fig.~\ref{fig:remnant_errors} shows the joint
distribution of kick magnitude and kick direction error for \RemnantModel,
showing that direction errors are larger at low kick magnitudes.  Our error in
kick direction is below $\roughly 0.2$ radians whenever $v_f\gtrsim
2\times10^{-3} c$.

We also compare the performance of our fits against several existing fitting
formulae for remnant mass, spin, and kick which we denote as follows: HBMR
(\cite{Hofmann:2016yih, Barausse:2012qz} with $n_M\!=\!n_J\!=\!3$), UIB
\cite{Jimenez-Forteza:2016oae}, HL \cite{Healy:2016lce}, HLZ
\cite{Healy:2014yta}, and CLZM (\cite{Gonzalez:2006md, Campanelli:2007ew,
Lousto:2007db, Lousto:2012su, Lousto:2012gt} as summarized in
\cite{Gerosa:2016sys}). To partially account for spin precession, these fits
are corrected as described in Ref.~\cite{mcdaniel2016} and used in current
LIGO/Virgo analyses \cite{TheLIGOScientific:2016pea, Abbott:2017vtc}: spins are
evolved using PN from relaxation to the Schwarzschild innermost stable circular
orbit, and final UIB and HL spins are post-processed by adding the sum of the
in-plane spins in quadrature. Figure \ref{fig:remnant_errors} shows that our
procedure to predict remnant mass, spin magnitude, and kick magnitude for
precessing systems is more accurate than these existing fits by at least an
order of magnitude.

Our fits appear to outperform the NR simulations when estimating the spin
direction. Once again, this is due to the post-junk-radiation initial spins of
the two highest resolutions being inconsistent with each other for some of our
simulations, so that different resolutions represent different physical systems
(cf. App. \ref{sec:PBandJ}). Therefore, the errors estimated by comparing the
two highest resolutions is a poor estimate of the actual truncation error for
these cases.  The actual truncation error is likely to be close to the errors
reproduced by the surrogate.

\begin{figure}[thb]
\includegraphics[width=0.45\textwidth]{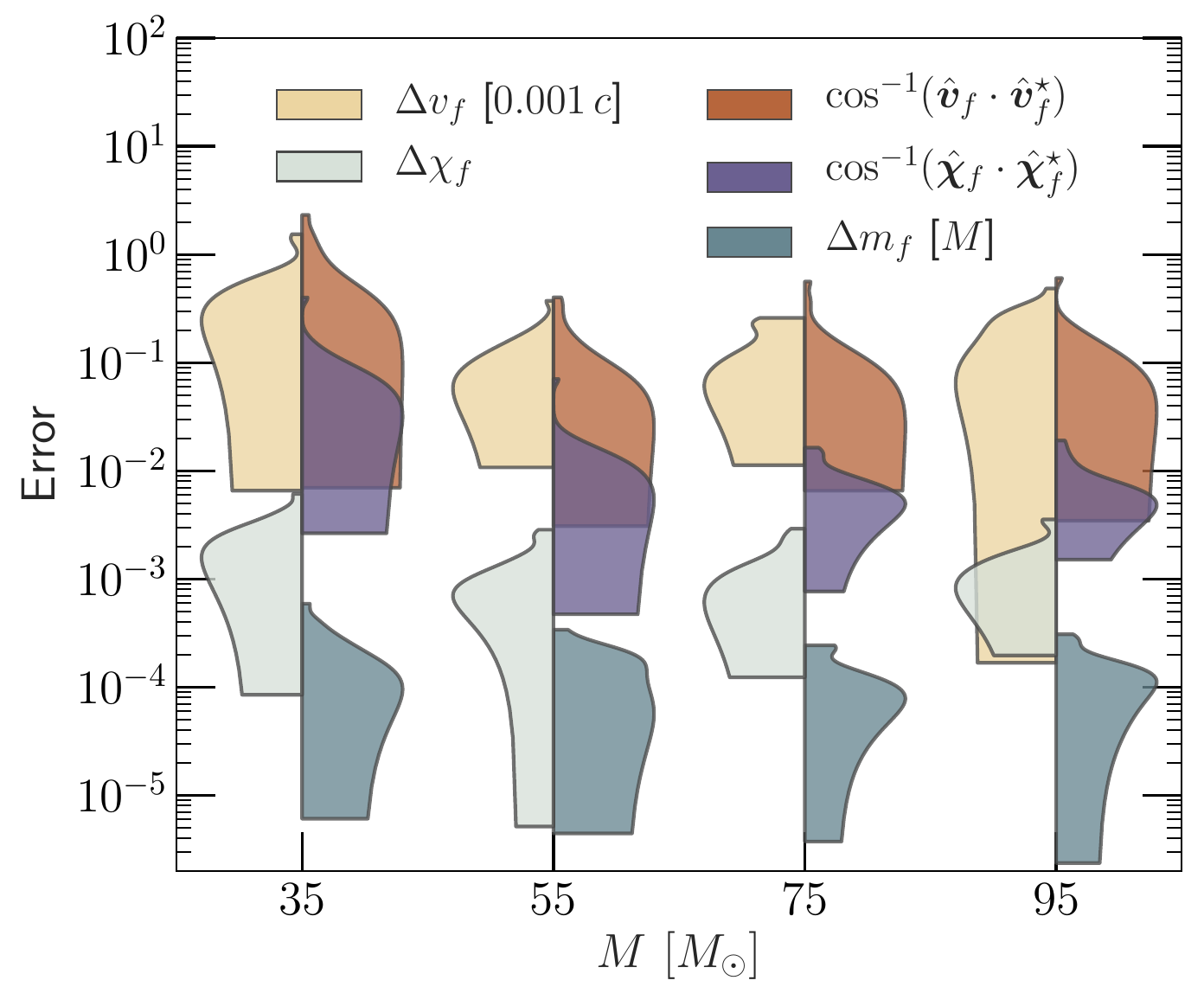}
\caption{Errors for \RemnantModel in predicting remnant properties when
    spins are specified at an orbital frequency of $f_{0}\!=\!10 \,{\rm Hz}$.
    For four different total masses, we compute the differences between the
    surrogate prediction of various remnant properties with the value obtained
    in the NR simulation.  For each mass, these differences are shown as a
    vertical histogram.  Note that the distributions in these plots are
    normalized to have a fixed height, not fixed area.
}
\label{fig:remnant_longNR}
\end{figure}

The \RemnantModel fits in Fig.~\ref{fig:remnant_errors} are evaluated using the
NR spins at $\tmHundred$ as inputs. In typical applications, one may have
access to the spins only at the start of the waveform, rather than at
$\tmHundred$.  For this case, as described in Sec.~\ref{sec:remnant_sur}, we
use a combination of PN and NRSur7dq4 to evolve the spins from any given
starting frequency to $\tmHundred$. These spins are then used to evaluate the
\RemnantModel fits.  Thus, spins can be specified at any given orbital frequency
and are evolved consistently before estimating the final BH properties.  This
is a crucial improvement (introduced by Ref.~\cite{Varma:2018aht}) over
previous results, which, being calibrated solely to non-precessing systems,
suffer from ambiguities regarding the time/frequency at which spins are
defined.

Figure~\ref{fig:remnant_longNR} shows the errors in \RemnantModel when the spins
are specified at an orbital frequency $f_0\!=\!10\, {\rm Hz}$.  These errors
are computed by comparing against 23 long NR ($3\times10^4M$ to $10^5M$ in
length) simulations~\cite{SXSCatalog2018} with mass ratios $q\leq4$ and
generically oriented spins with magnitudes $\chi_1, \chi_2 \sim 0.5$. None of
these simulations were used to train the fits. Longer PN evolutions are needed
at lower total masses, and the errors are therefore larger. These errors will
decrease with an improved spin evolution procedure.  Note, however, that our
predictions are still more accurate than those of existing fitting formulae
(cf.~Fig.~\ref{fig:remnant_errors}).

\section{Conclusion}
\label{sec:conclusion}

We present new NR surrogate models for precessing BBH systems with generic
spins and unequal masses. In particular, we model the two most-used outputs of
NR simulations: the gravitational waveform and the properties (mass, spin, and
recoil kick) of the final BH formed after the merger. Trained against 1528 NR
simulations with mass ratios $q\leq4$, spin magnitudes $\chi_{1,2} \leq 0.8$,
and generic spin directions, both these models are shown to reproduce the NR
simulations with accuracies comparable to those of the simulations themselves.

The waveform model, NRSur7dq4, includes all spin-weighted spherical harmonic
modes up to $\ell=4$. The precession frame dynamics and spin evolution of the
BHs are also modeled as byproducts. Through a cross-validation study, we show
that the mismatches for NRSur7dq4 against NR computed with the Advanced LIGO
design sensitivity noise curve are always $\lesssim8\times10^{-3}$ at the $95$
percentile level over the mass range $50-200 M_{\odot}$. This is at least an
order of magnitude improvement over existing waveform models. NRSur7dq4 is made
publicly available through the gwsurrogate~\cite{gwsurrogate} Python package,
with example evaluation code at Ref.~\cite{SpECSurrogates}.

For the final BH model, \RemnantModel, the  $95$th percentile errors are
$\roughly 5 \into 10^{-4} M$ for mass, $\roughly 2 \into 10^{-3}$ for spin
magnitude, $\roughly 4 \into 10^{-4}\,c$ for kick magnitude. Once again, these
are lower than that of existing models by at least an order of magnitude.  In
addition, we also model the spin and kick directions. Moreover, the GPR
methods employed here naturally provide error estimates along with the fitted
values. These uncertainty estimates can be incorporated into data analysis
applications to marginalize over systematic uncertainties.  \RemnantModel is
made publicly available through the surfinBH~\cite{surfinBH} Python package,
which includes an example evaluation code. 

Futher, we provide a Python package, binaryBHexp~\cite{binaryBHexp}, to
visualize the complex precessing dynamics as predicted by these surrogate
models~\cite{Varma:2018rcg}.

\subsection{Future work}

In App.~\ref{sec:extrap} we test the performance of these surrogate models when
extrapolated outside their training range to $q=6$. We find that our models
become worse at these mass ratios, but are still comparable or better than
existing models.  Unfortunately, suitable precessing simulations are currently
not available for testing at intermediate mass ratios $4<q<6$. In general, we
advice caution with extrapolation. A natural improvement of both NRSur7dq4 and
\RemnantModel is to extend their range of validity with new training
simulations at higher mass ratios and spin magnitudes. We note, however, that
both these regimes are increasingly expensive to model in NR.

\begin{figure}[thb]
\includegraphics[width=0.5\textwidth]{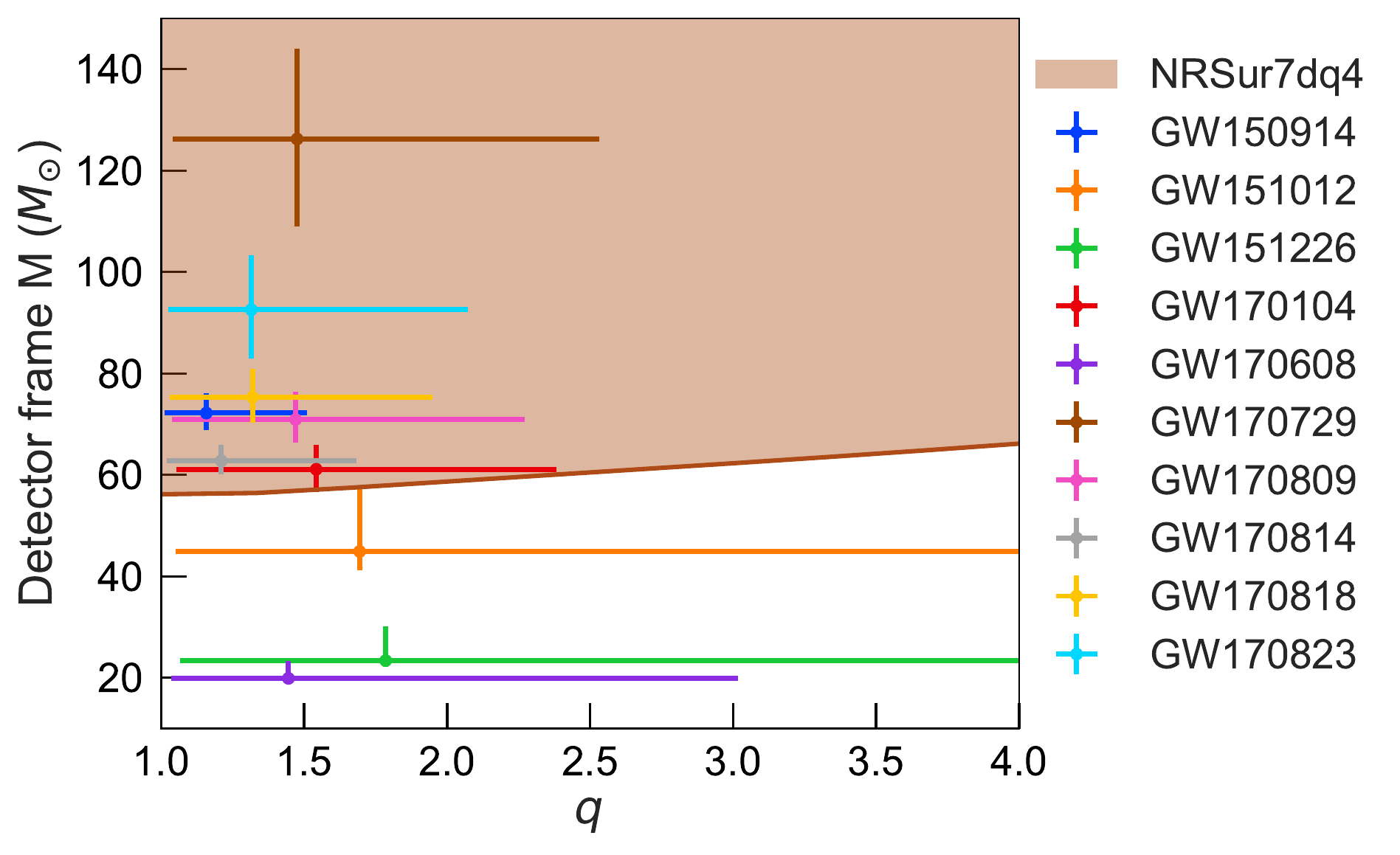}
\caption{The shaded region shows the regime of validity of the (2,2) mode of
    NRSur7dq4 with a starting frequency of $20$ Hz. Also shown are the
    parameter ranges for the 10 BBH signals seen by LIGO and Virgo during the
    first two observing runs~\cite{LIGOScientific:2018mvr}. The markers
    indicate the median values of the marginalized posteriors for the detector
    frame total mass M and mass ratio q. The error bars indicate the range
    between the 5th percentile and 95th percentile values of the posteriors.
}
\label{fig:mass_range}
\end{figure}

Another important limitation of these models is that they are restricted to the
same length as the NR simulations (starting time of $\sim4300M$ before the peak
or about 20 orbits). For LIGO, assuming a starting GW frequency of 20 Hz, the
(2, 2) mode of the surrogate is valid for total masses $M\gtrsim66M_{\odot}$.
This number, however, depends on the mass ratio. Fig.~\ref{fig:mass_range}
shows the mass range of validity of NRSur7dq4 as a function of mass ratio. We
compare this with the parameters of the 10 BBH detections seen by LIGO and
Virgo in the first two observing runs~\cite{LIGOScientific:2018mvr}.  NRSur7dq4
sufficiently covers the posterior spread of most but not all of these
detections, the main limitation being the number of orbits covered by the
model. However, see Ref.~\cite{Kumar:2018hml} for an example of NR surrogates
used in data analysis with GW signals.

A promising avenue to extend the length of the waveforms is to ``hybridize''
the simulations using PN waveforms in the early inspiral. This approach already
was found to be successful for the case of aligned-spin BBH
\cite{Varma:2018mmi}, but still needs to be generalized to precessing spins.
Furthermore, it is not clear if the current length of the NR simulations is
sufficient to guarantee good attachment of the PN and NR waveforms for
precessing BBH.

Despite these limitations, in their regime of validity, the models presented in
the paper are the most accurate models currently available for precessing BBHs.
As shown in this paper, our models rival the accuracy of the NR simulations,
while being very cheap to evaluate. As more and more BBHs are detected at
higher signal-to-noise ratios, fast yet accurate models such as these will
contribute to turning GW astronomy into high precision science.

\begin{acknowledgments}
We thank Dan Hemberger, Kevin Barkett, Marissa Walker, Matt Giesler, Nils
Deppe, Francois Hebert, Maria Okounkova, and Geoffrey Lovelace for helping
carry out the new SpEC simulations used in this work. V.V. and M.S. are
supported by the Sherman Fairchild Foundation, and NSF grants PHY--170212 and
PHY--1708213 at Caltech.  L.E.K.  acknowledges support from the Sherman
Fairchild Foundation and NSF grant PHY-1606654 at Cornell.  S.E.F is partially
supported by NSF grant PHY-1806665.  This work used the Extreme Science and
Engineering Discovery Environment (XSEDE), which is supported by National
Science Foundation grant number ACI-1548562.  This research is part of the Blue
Waters sustained-petascale computing project, which is supported by the
National Science Foundation (awards OCI-0725070 and ACI-1238993) and the state
of Illinois. Blue Waters is a joint effort of the University of Illinois at
Urbana-Champaign and its National Center for Supercomputing Applications.
Simulations were performed on NSF/NCSA Blue Waters under allocation NSF
PRAC--1713694; on the Wheeler cluster at Caltech, which is supported by the
Sherman Fairchild Foundation and by Caltech; and on XSEDE resources Bridges at
the Pittsburgh Supercomputing Center, Comet at the San Diego Supercomputer
Center, and Stampede and Stampede2 at the Texas Advanced Computing Center,
through allocation TG-PHY990007N.  Computations for building the model were
performed on Wheeler.
\end{acknowledgments}

\appendix

\section{Evaluating surrogates at larger mass ratios}
\label{sec:extrap}

\begin{figure}[thb]
\includegraphics[width=0.45\textwidth]{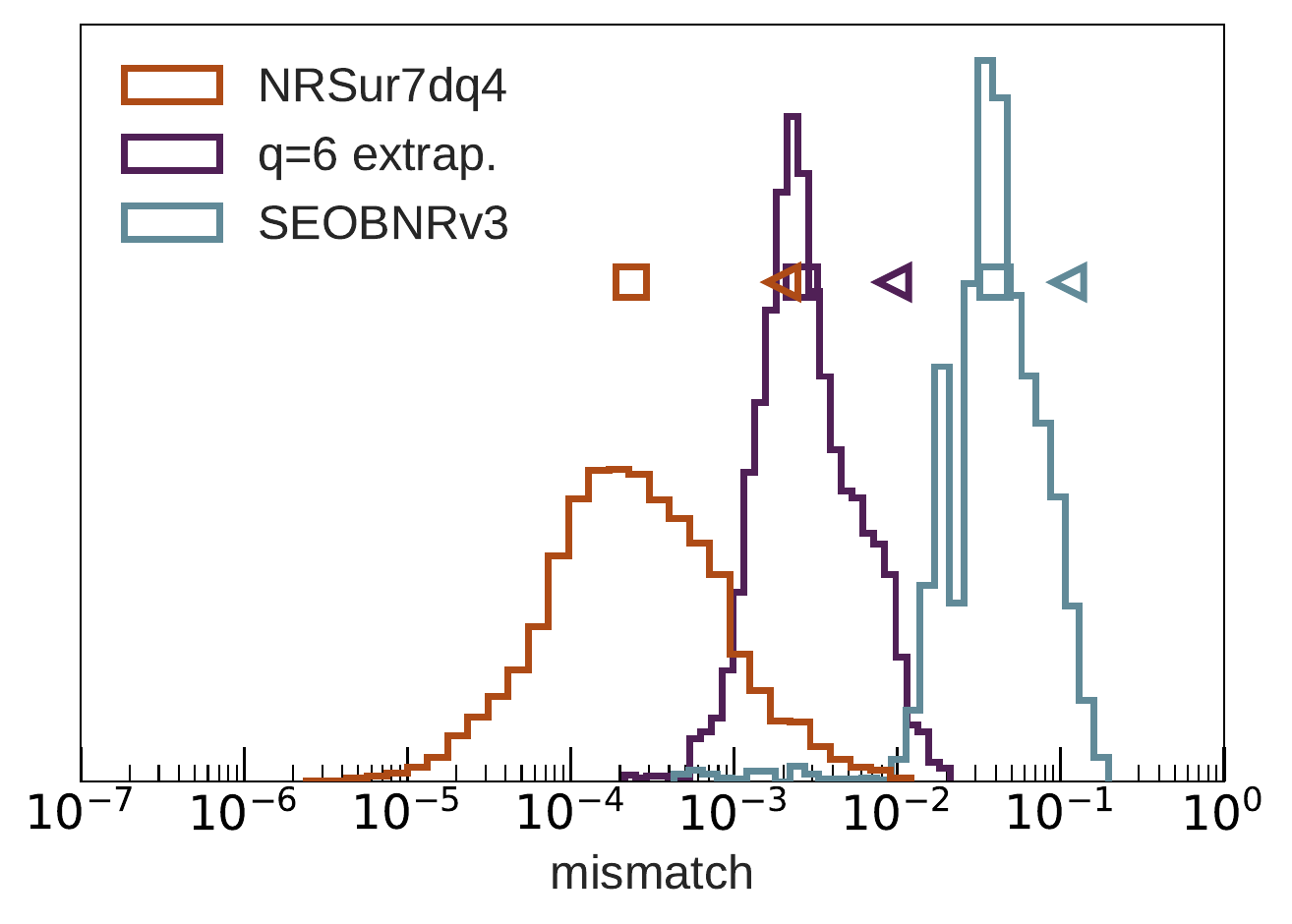}
\caption{Mismatch histogram when extrapolating the NRSur7dq4 waveform model to
    mass ratio $q=6$. Also shown are mismatches for \eobmodel. The mismatches
    are computed using a flat noise curve. The training range errors from
    Fig.~\ref{fig:waveform_errors_flat} are reproduced here for comparison. The
    square (triangle) markers indicate median ($95$th percentile) values.
}
\label{fig:waveform_extrap}
\end{figure}

\begin{figure}[thb]
\includegraphics[width=0.45\textwidth]{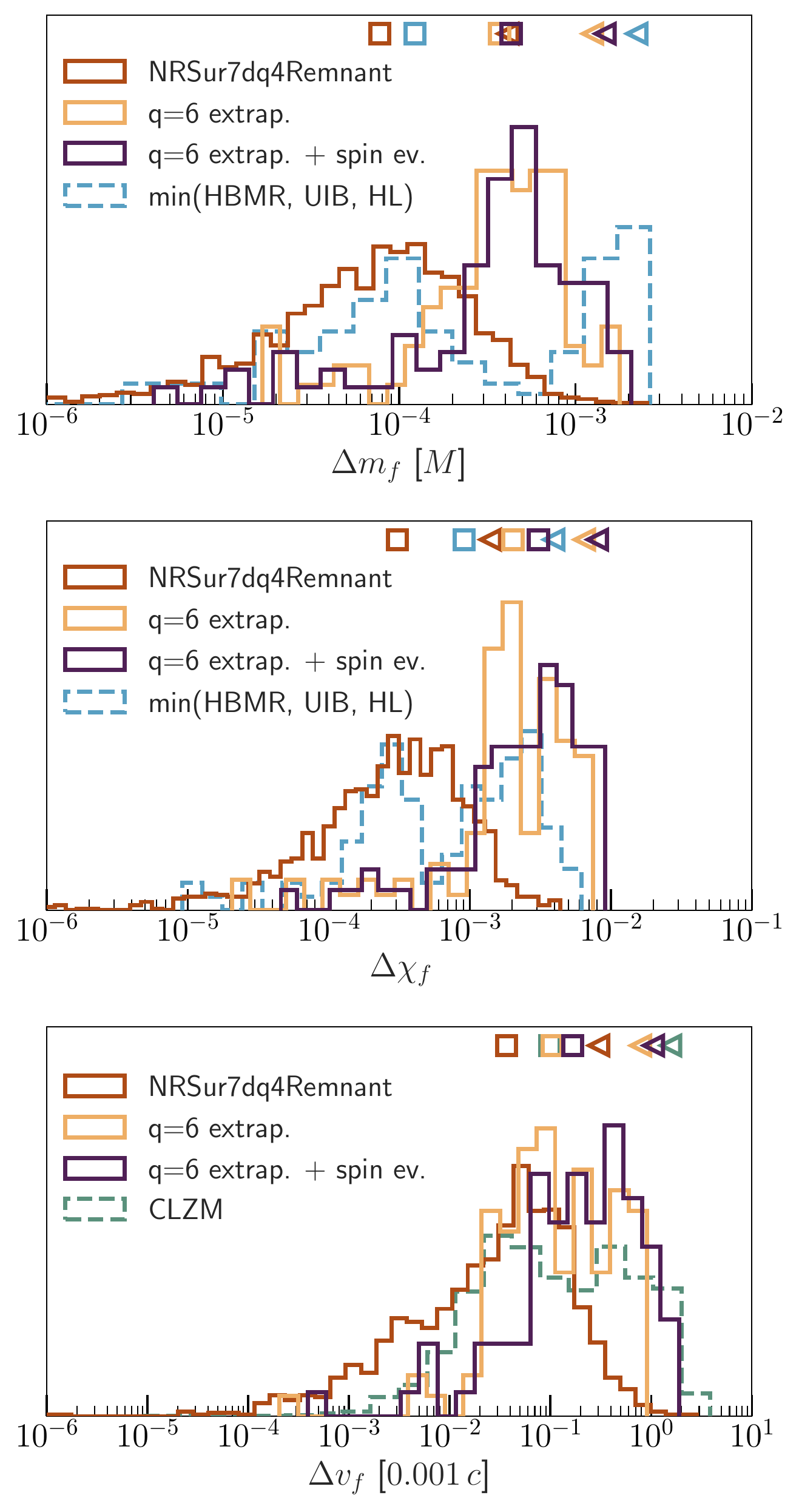}
\caption{Error histograms of the remnant mass, spin magnitude and kick
    magnitude when extrapolating \RemnantModel to mass ratio $q=6$. The
    training range errors from Fig.~\ref{fig:remnant_errors} are reproduced
    here for comparison.  We show errors using the NR spins at $\tmHundred$
    (yellow) as well as the initial NR spins (blue) as inputs for the model.
    Also shown are the errors for existing fitting formulae described in
    Sec.~\ref{sec:remnant_errs}; for the final mass and spin, we only show the
    minimum error among the HBMR, UIB, and HL fits. The square (triangle)
    markers indicate median ($95$th percentile) values.
}
\label{fig:remnant_extrap}
\end{figure}

In this Appendix we assess the performance of the NRSur7dq4 and \RemnantModel
models when evaluated at mass ratio $q=6$. Doing so is effectively an
extrapolation because $q=6$ is outside the training range of the surrogates
($q\leq4$). The surrogate models are compared against 100 NR simulations with
$q=6$ and generically precessing spins with magnitudes $\chi_1, \chi_2 \leq
0.8$. These simulations have been assigned the identifiers SXS:BBH:2164 -
SXS:BBH:2263, and are made publicly available through the SXS public
catalog~\cite{SXSCatalog}.

Figure~\ref{fig:waveform_extrap} shows the $q=6$ extrapolation mismatches for
NRSur7dq4. Also shown are the mismatches for \eobmodel when compared against
the same simulations. The mismatches are computed in the same manner as in
Fig.~\ref{fig:waveform_errors_flat}, which we reproduce here for comparison.
The surrogate errors become noticeably worse when extrapolating to $q=6$, but
are still much smaller than the corresponding errors for \eobmodel.

Fig.~\ref{fig:remnant_extrap} shows the performance of \RemnantModel when
extrapolating to $q=6$. We show the errors when the fits are evaluated using
the NR spins at $\tmHundred$ as well as when the spins are specified at the
start of the NR simulations. In the latter case, we use the extrapolated
dynamics surrogate of NRSur7dq4 to evolve the spins to $\tmHundred$ and then
evaluate the fits. We reproduce the training range errors from
Fig.~\ref{fig:remnant_errors} for comparison. Also shown are the errors for the
existing fitting formulae described in Sec.~\ref{sec:remnant_errs} when
compared against the same simulations. We find that \RemnantModel performs
noticeably worse when extrapolated to $q=6$ but is still slightly better than
the existing fitting formulae, except for the final spin where the existing
fitting formulae perform slightly better.

In general, we find that the NRSur7dq4 and \RemnantModel models become worse
with extrapolation to $q=6$ but are still better or comparable to existing
models.  Unfortunately, we do not have enough suitable precessing simulations
with $4<q<6$ with which to test at what mass ratio the degradation of these
surrogate models becomes significant. We leave these tests, as well as
extending the models to larger mass ratios by adding NR simulations, to future
work.

\section{On the high mismatch tail in NR errors}
\label{sec:PBandJ}

\begin{figure}[thb]
\includegraphics[width=0.5\textwidth]{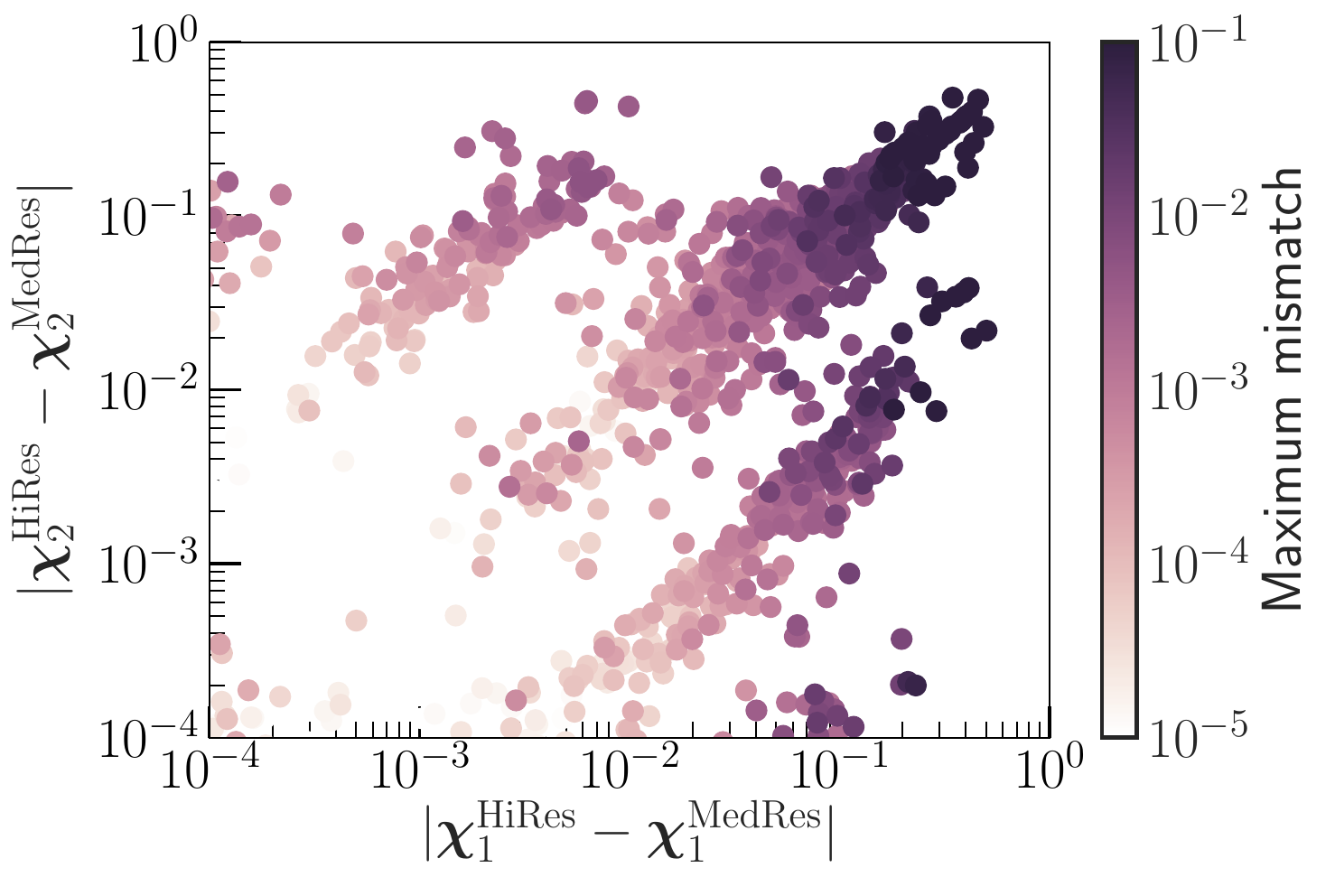}
\caption{Dependence of the NR resolution error on the difference in the
    relaxation-time spins of the two highest resolutions (labeled HiRes and
    MedRes). The horizontal (vertical) axis shows the difference between the
    spin of the heavier (lighter) BH. The colors show the largest (flat noise)
    mismatch between the waveforms of the two resolutions over different points
    in the sky. Large mismatches occur when the difference between the
    relaxation-time spins of the two resolutions is large.
}
\label{fig:NRmismatch_vs_LevErr}
\end{figure}

The histogram of NR errors in Fig.~\ref{fig:waveform_errors_flat} shows a
significant tail to the right, i.e. at large mismatches.  In
Sec.~\ref{sec:waveform_errs}, this tail was attributed to different resolutions
of the same NR simulation having different physical parameters, namely the
``initial'' spins, which are measured at the relaxation
time~\cite{SXSCatalog2018} after the poorly-resolved junk-radiation transients
have settled.  In this Appendix we provide some evidence for this claim.
Figure~\ref{fig:NRmismatch_vs_LevErr} shows the maximum mismatch (with a flat
noise curve) over points in the sky versus the difference in the
relaxation-time dimensionless spins between the two highest resolutions. We
refer to the two highest resolutions as HiRes and MedRes, and their
corresponding relaxation-time dimensionless spins are denoted by
($\bchi^{\mathrm{HiRes}}_{1}, \bchi^{\mathrm{HiRes}}_{2}$) and
($\bchi^{\mathrm{MedRes}}_{1}, \bchi^{\mathrm{MedRes}}_{2}$), respectively.  We
note that the largest mismatch occurs when the spin difference is largest
between the two resolutions. For a significant fraction of the simulations the
spins can be different by about $0.1$; for these cases the two resolutions
essentially represent two different physical systems, so the difference in
waveforms between the two resolutions fails to be a good estimate of the
truncation error in the simulations.

\begin{figure}[tbh]
\includegraphics[width=0.45\textwidth]{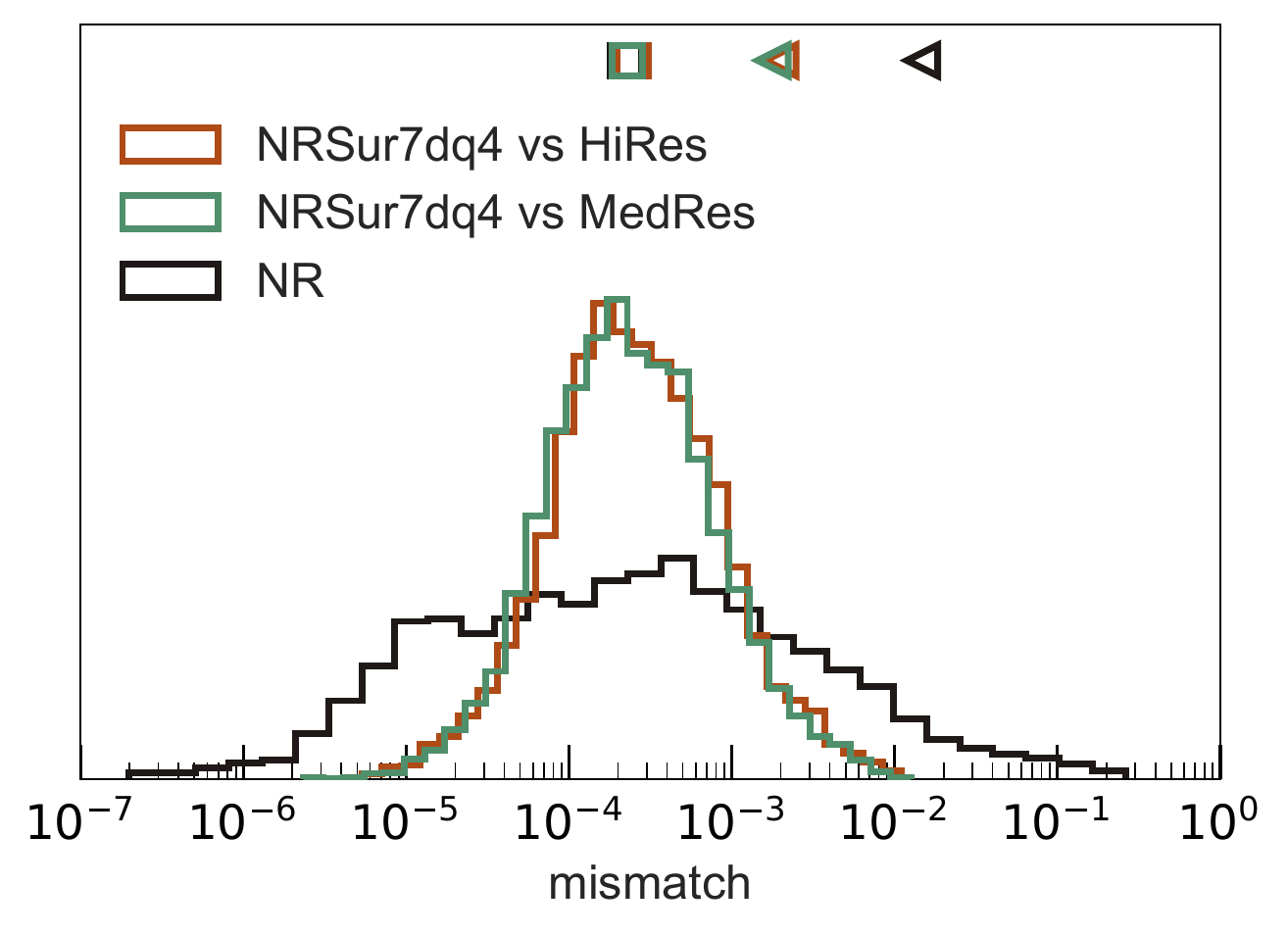}
\caption{Mismatch histograms for NRSur7dq4 when compared against the two
    highest available NR simulations (referred to as HiRes and MedRes). Also
    shown are mismatches between the two resolutions (labeled NR). The
    ``NRSur7dq4 vs HiRes'' and NR errors are the same as the red and black
    histograms, respectively, in Fig.~\ref{fig:waveform_errors_flat}. These are
    flat noise mismatches, computed at several points in the sky. The square
    (triangle) markers indicate median ($95$th percentile) values.
}
\label{fig:waveform_errors_Lev2}
\end{figure}

Figure~\ref{fig:NRmismatch_vs_LevErr} suggests that the high NR mismatch tail
of Fig.~\ref{fig:waveform_errors_flat} is artificially large, and if the two
resolutions were to correspond to the same physical system, the tail would be
shorter.  We test this in Fig.~\ref{fig:waveform_errors_Lev2}, where we compare
the surrogate against the MedRes simulations, but use the spins of the MedRes
simulation ($\bchi^{\mathrm{MedRes}}_{1}, \bchi^{\mathrm{MedRes}}_{2}$) to
evaluate the surrogate. The surrogate mismatches against the HiRes simulations
as well as the NR resolution mismatches (HiRes vs MedRes) are reproduced from
Fig.~\ref{fig:waveform_errors_flat} for comparison.  We note that the surrogate
mismatches when compared against the MedRes simulations always lie below
$\sim10^{-2}$ and do not have the high mismatch tail seen for the NR resolution
mismatches. In this test, we are treating the surrogate, which is trained on
the HiRes simulations, as a proxy for the HiRes dataset. Evaluating the
surrogate with the parameters of a MedRes simulation is treated as a proxy for
performing the HiRes simulation with the same parameters. Therefore, the green
histogram in Fig.~\ref{fig:waveform_errors_Lev2} can be treated as the ``true''
resolution error when the parameters of the resolutions are the same. As
expected for this case, this estimate of the resolution error agrees with the
errors for the surrogate model (red histogram).

\begin{figure}[bth]
\includegraphics[width=0.45\textwidth]{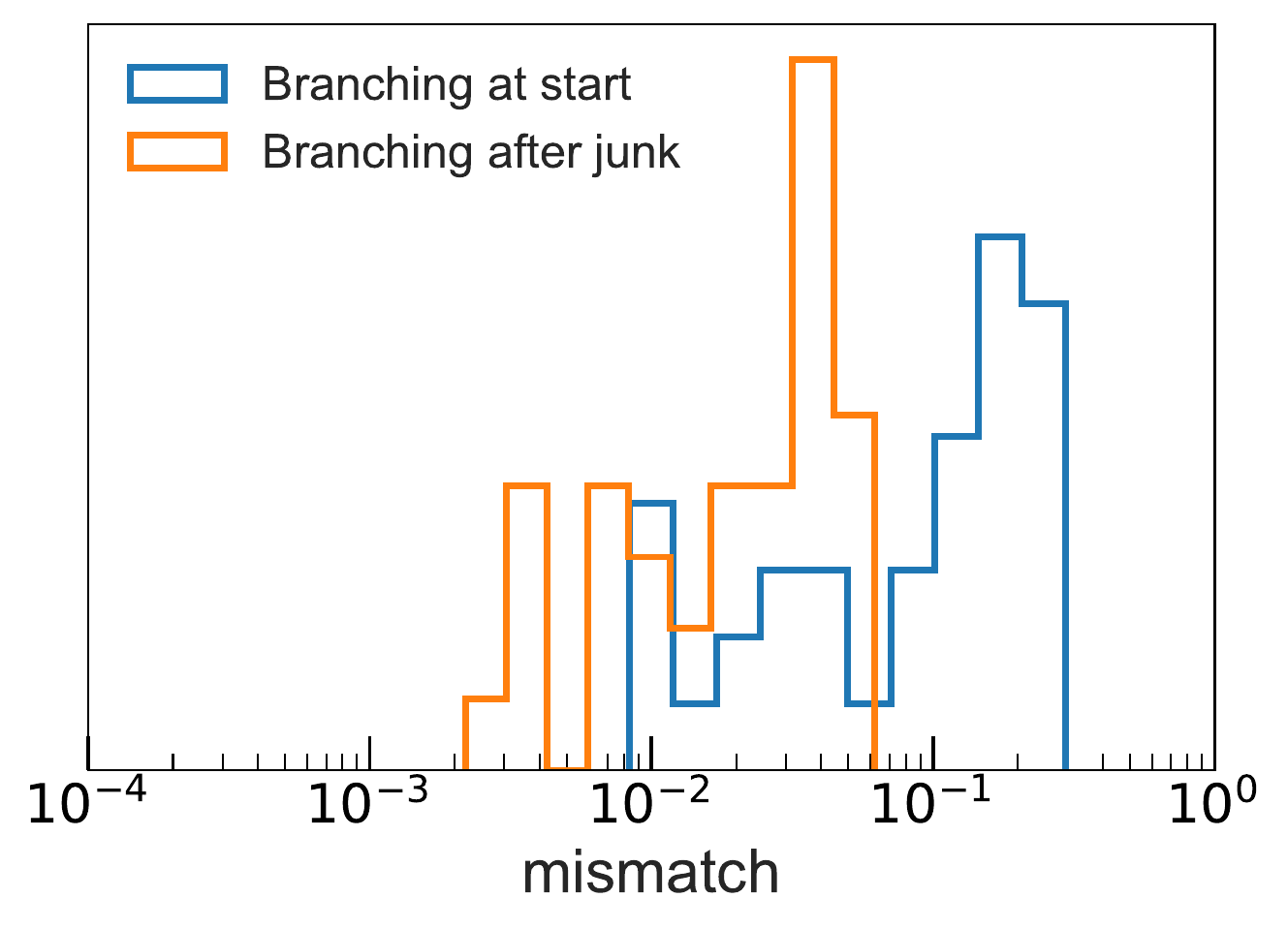}
\caption{NR resolution mismatches for the simulation leading to the largest NR
    mismatch in Fig.~\ref{fig:waveform_errors_flat}.  The different samples in
    the histogram correspond to comparisons at different angles on the sky. The
    blue histogram shows the current resolution errors when the two resolutions
    start with the same initial data at the start of the simulation. All points
    in the blue histogram are the same as those included in
    Fig.~\ref{fig:waveform_errors_flat}.  The green histogram shows the
    resolution errors for the same case when the two resolutions start with the
    same initial data at $\sim1000M$ after start, at which point the junk
    radiation has left the simulation domain.
}
\label{fig:PBandJ}
\end{figure}

Together, Figs.~\ref{fig:NRmismatch_vs_LevErr} and
\ref{fig:waveform_errors_Lev2} show that the high NR mismatch tail in
Fig.~\ref{fig:waveform_errors_flat} is due to the difference in the parameters
of the different NR resolutions.  We believe this difference arises from
spurious initial transients known as ``junk radiation''. These transients
result from initial data that do not precisely represent a snapshot of a binary
that has evolved from $t=-\infty$. The transients quickly leave the simulation
domain after about one or two binary orbits.  It is computationally expensive
to resolve the high spatial and temporal frequencies of the transients, so we
typically choose not to resolve these transients at all, and instead we simply
discard the initial part of the waveform. Because some of the transients carry
energy and angular momentum down the BHs, the masses and spins are modified, so
we measure ``initial'' masses and spins at a relaxation time
\cite{SXSCatalog2018} deemed sufficiently late that the transients have decayed
away. Because we do not fully resolve the transients, their effect on the
masses and spins are not always convergent with resolution.

This issue should ideally be resolved with improved, junk-free initial data
(see Ref.~\cite{Varma:2018sqd} for steps in this direction). In the meantime,
we propose a change in how SpEC performs different resolutions for the same
simulation. Currently, initial data are constructed by solving the Einstein
constraint equations~\cite{Lovelace:2008tw, Ossokine:2015yla}. The same
constraint-satisfying initial data are then interpolated onto several grids of
different resolution, and Einstein's equations are evolved on each grid
independently.  Our proposal is to first evolve the initial data using the high
resolution grid until the transients leave the simulation domain, and then
interpolate the data at that time onto grids of lower resolution, and evolve
Einstein's equations on these lower-resolution grids independently.  This way
all resolutions start with the same initial data at a time after transients
have decayed away instead of at the start of the simulation, and the masses and
spins of the black holes should be convergent.

This proposal is tested in Fig.~\ref{fig:PBandJ} for the case leading to the
largest NR mismatch in Fig.~\ref{fig:waveform_errors_flat}. We perform the
resolution branching at $t\sim1000M$ after the start of the high resolution
simulation.  The outer boundary is at $\sim600M$ and this is sufficient time
for junk radiation to leave the simulation domain. We find that the mismatches
decrease significantly when the resolution branching is done post-junk, as the
resolutions now correspond to the same physical system.

\section*{References}
\bibliography{References}

\end{document}